\documentclass[prd,twocolumn,showpacs,superscriptaddress,nofootinbib]{revtex4}

\usepackage{amsfonts}
\usepackage{amsmath}
\usepackage{amssymb}
\usepackage{upgreek}
\usepackage{bm}
\usepackage{dcolumn}
\usepackage{epsfig}
\usepackage{graphicx}
\usepackage{graphics}
\usepackage[latin1]{inputenc}
\usepackage{latexsym}
\usepackage{rotating}
\usepackage{hyperref}
\usepackage{xspace} % Sensible space treatment at end of simple macros
\usepackage[usenames,dvipsnames]{color}

\usepackage{ulem}
\definecolor {darkgreen}{rgb}{0.2,0.7,0.2}

\newcommand\be{\begin{equation}}
\newcommand\ba{\begin{eqnarray}}
\newcommand\ee{\end{equation}}
\newcommand\ea{\end{eqnarray}}

\newcommand{\GR}{{\mbox{\tiny GR}}}
\newcommand{\ppE}{{\mbox{\tiny ppE}}}

\newcommand{\beq}{\begin{equation}}
\newcommand{\eeq}{\end{equation}}
\newcommand{\bes}{\begin{subequations}}
\newcommand{\ees}{\end{subequations}}
\newcommand{\beqn}{\begin{eqnarray*}}
\newcommand{\eeqn}{\end{eqnarray*}}

\newcommand{\f}[2]{\frac{#1}{#2}}

\begin{document}
\title{Constraining Lorentz-Violating, Modified Dispersion Relations \\ with Gravitational Waves}
%Constraining Generic Lorentz Violation and the Speed of the Graviton with Gravitational Waves

\author{Saeed Mirshekari}
\affiliation{McDonnell Center for the Space Sciences, Department of Physics,
Washington University, St. Louis MO 63130 USA}

\author{Nicol\'as Yunes}
\affiliation{MIT and Kavli Institute, Cambridge, MA 02139, USA.}
\affiliation{Department of Physics, Montana State University, Bozeman, MT 59717, USA.}

\author{Clifford M.~Will}
\affiliation{McDonnell Center for the Space Sciences, Department of Physics,
Washington University, St. Louis MO 63130 USA}

\date{October 2011}

%%%%%%%%%%%%%%%%%%%%%%%%%%%%%%%%%%%%%%%%%%%%%%%%%
\begin{abstract} 

Modified gravity theories generically predict a violation of Lorentz invariance, which may lead to a modified dispersion relation for propagating modes of gravitational waves. We construct a parametrized dispersion relation that can reproduce a range of known Lorentz-violating predictions and investigate their impact on the propagation of gravitational waves. A modified dispersion relation forces different wavelengths of the gravitational wave train to travel at slightly different velocities, leading to a modified phase evolution observed at a gravitational-wave detector. We show how such corrections map to the waveform observable and to the parametrized post-Einsteinian framework, proposed to model a range of deviations from General Relativity. Given a 
gravitational-wave detection, the lack of evidence for such corrections could then be used to place a constraint on Lorentz violation.  The constraints we obtain are tightest for dispersion relations that scale with small power of the graviton's momentum and deteriorate for a steeper scaling. 

\end{abstract}


\maketitle

%%%%%%%%%%%%%%%%%%%%%%%%%%%%%
%%%%%%%%%%%%%%%%%%%%%%%%%%%%%
\section{Introduction}

After a century of experimental success, Einstein's fundamental theories, ie.~the special theory of relativity and the General theory of Relativity (GR), are beginning to be questioned. As an example, consider the observation of ultra-high-energy cosmic rays. In relativity, there is a threshold of $\sim 5\times 10^{19} \; {\rm{eV}}$ (GZK limit) for the amount of energy that charged particles can carry, while cosmic rays have been detected with higher energies~\cite{Bird:1994uy}. On the theoretical front, theories of quantum gravity also generically predict a deviation from Einstein's theory at sufficiently large energies or small scales. In particular, Lorentz violation seems ubiquitous in such theories.  These considerations motivate us to study the effects of Lorentz violation on gravitational wave observables.

Einstein's theory will soon be put to the test through a new type of observation: gravitational waves (GWs). Such waves are (far-field) oscillations of spacetime that encode invaluable and detailed information about the source that produced them. For example, the inspiral, merger and ringdown of compact objects (black holes or neutron stars) are expected to produce detectable waves that will access horizon-scale curvatures and energies. Gravitational waves may thus provide new hints as to whether Einstein's theory remains valid in this previously untested regime. 

Gravitational-wave detectors are today a reality.  Ground-based interferometers, such as the Advanced Laser Interferometer Gravitational Observatory (Ad.~LIGO)~\cite{ligo,Abbott:2007kv,2010CQGra..27h4006H} and Advanced~Virgo~\cite{virgo}, are currently being updated, and are scheduled to begin data acquisition by 2015. Second generation detectors, such as the Einstein Telescope (ET)~\cite{et,Punturo:2010zz} and the Laser Interferometer Space Antenna (LISA)~\cite{lisa,Prince:2003aa}, are also being planned for the next decade. Recent budgetary constraints in the United States have cast doubt on the status of LISA, but the European Space Agency is still considering a descoped, LISA-like mission (an NGO, or New Gravitational Observatory).  The detection of gravitational waves is, of course, not a certainty, as the astrophysical event rate is highly uncertain. However, there is consensus that advanced ground detectors should observe a few gravitational-wave events by the end of this decade.  

Some alternative gravity theories endow the graviton with a mass~\cite{Will:1993ns}.  Massive gravitons would travel slower than the speed of light, but most importantly, their speed would depend on their energy or wavelength. Since gravitational waves emitted by compact binary inspirals chirp in frequency, gravitons emitted in the early inspiral will travel more slowly than those emitted close to merger, leading to a frequency-dependent gravitational-wave dephasing, compared to the phasing of a massless general relativistic graviton. If such a dephasing is not observed, then one could place a constraint on the graviton mass~\cite{Will:1997bb}.
A Lorentz-violating graviton dispersion relation leaves an additional
imprint on the propagation of gravitational waves, irrespective of the generation mechanism.  Thus a bound on the dephasing effect could also bound the degree of Lorentz violation.

In this paper, we construct a framework to study the impact of a Lorentz-violating dispersion relation on the propagation of gravitational waves. We begin by proposing a generic, but quantum-gravitational inspired, modified dispersion relation, given by
\be
\label{dispersion}
E^2=p^2c^2 + m_{g}^2c^4 + \mathbb{A} p^\alpha c^{\alpha} ,
\ee
where $m_g$ is the mass of the graviton and $\mathbb{A}$ and $\alpha$ are two Lorentz-violating parameters that characterize the GR deviation ($\alpha$ is dimensionless while $\mathbb{A}$ has dimensions of $[{\rm{energy}}]^{2-\alpha}$).  We will assume that $\mathbb{A}/(cp)^{2-\alpha} \ll 1$.
When either $\mathbb{A} = 0$ or $\alpha = 0$, the modification reduces to that of a massive graviton. When $\alpha=(3,4)$, one recovers predictions of certain quantum-gravitation inspired models. This modified dispersion relation introduces Lorentz-violating deviations in a continuous way, such that when the parameter $\mathbb{A}$ is taken to zero, the dispersion relation reduces to that of a simple massive graviton.

The dispersion relation of Eq.\ (\ref{dispersion}) modifies the gravitational waveform observed at a detector by correcting the phase with certain frequency-dependent terms. In the stationary-phase approximation (SPA), the Fourier transform of the waveform is corrected by a term of the form $\zeta(\mathbb{A}) u^{\alpha-1}$, where $u = \pi {\cal{M}} f$ is a dimensionless measure of the gravitational-wave frequency with ${\cal{M}}$ the so-called ``chirp mass''. We show that such a modification can be easily mapped to the recently proposed parametrized post-Einsteinian framework (ppE)~\cite{Yunes:2009ke,Cornish:2011ys} for an appropriate choice of ppE parameters. 

In deriving the gravitational-wave Fourier transform we must assume a functional
form for the waveform as emitted at the source so as to relate the time of arrival at the detector to the gravitational-wave
frequency.  In principle, this would require a prediction for the equations of motion and gravitational-wave emission for each Lorentz violating theory under study.  However few such theories have reached a sufficient state of development to produce such predictions.  On the other hand, it is reasonable to assume that the predictions will be not too different from those of general relativity.  For example, we argued~\cite{Will:1997bb} that for a theory with a massive graviton, the differences would be of order $(\lambda/\lambda_g)^2$, where $\lambda$ is the gravitational wavelength, and $\lambda_g$ is the graviton Compton wavelength, and $\lambda_g \gg \lambda$ for sources of interest.  Similar behavior might be expected in Lorentz violating theories.   The important phenomenon is the accumulation of dephasing over the enormous propagation distances from source to detector, not the small differences in the source behavior.   As a result, we will use the standard general relativistic wave generation framework for the source waveform.

With this new waveform model, we then carry out a simplified (angle-averaged) Fisher-matrix analysis to estimate the accuracy to which the parameter $\zeta (\mathbb{A})$ could be constrained as a function of $\alpha$, given a gravitational-wave detection consistent with general relativity. We perform this study with a waveform model that represents a non-spinning, quasi-circular, compact binary inspiral, but that deviates from general relativity only through the effect of the modified dispersion relation on the propagation speed of the waves, via Eq.\ (\ref{dispersion}).  

To illustrate our results, we show in Table~\ref{summary} the accuracy to which Lorentz-violation in the $\alpha=3$ case could be constrained, as a function of system masses and detectors for fixed signal-to-noise ratio (SNR). The case $\alpha=3$ is a prediction of ``doubly special relativity''. The bounds on the graviton mass are consistent with previous studies~\cite{Will:1997bb,Will:2004xi,Berti:2004bd,Arun:2009pq,2010PhRvD..82l2001K,2010PhRvD..81f4008Y} (for a recent summary of current and proposed bounds on $m_g$ see~\cite{2011arXiv1107.3528B}).  The table here means that given a gravitational-wave detection consistent with GR, $m_{g}$ and ${\mathbb{A}}$ would have to be smaller than the numbers on the third and fourth columns respectively. 

%%%%%%%
\setlength{\tabcolsep}{5pt} 
\setlength{\extrarowheight}{1.5pt}
\begin{table}[ht]
\begin{tabular}{lcc| c c}
\hline\hline
Detector&$m_1$&$m_2$&					$m_g(eV)$		&${\mathbb A}(eV^{-1})$	\\
\hline
Ad.~LIGO&			1.4&1.4&$			3.71\times 10^{-22}$&	$7.36\times 10^{-8}$\\
${\rm{SNR}}=10$&		1.4&10&$				3.56\times 10^{-22}$&	$3.54\times 10^{-7}$\\
&					10&10&$				3.51\times 10^{-22}$&	$6.83\times 10^{-7}$\\
\hline
ET&					10&10&$				2.99\times 10^{-23}$&	$2.32\times 10^{-8}$\\
${\rm{SNR}}=50$&		10&100&$			4.81\times 10^{-23}$&	$1.12\times 10^{-6}$\\
&					100&100&$			6.67\times 10^{-23}$&	$3.34\times 10^{-6}$\\
\hline
NGO&				$10^4$&$10^4$&$		3.05\times 10^{-25}$&	$2.16\times 10^{-2}$\\
${\rm{SNR}}=100$&		$10^4$&$10^5$&$		2.46\times 10^{-25}$&	$0.147$\\
&					$10^5$&$10^5$&$		2.03\times 10^{-25}$&	$0.189$\\
&					$10^5$&$10^6$&$		2.09\times 10^{-25}$&	$9.57$\\
&					$10^6$&$10^6$&$		1.49\times 10^{-25}$&	$23.2$\\
\hline\hline
\end{tabular}
%%%%%%%
\caption{\label{summary} Accuracy to which graviton mass and the Lorentz-violating parameter ${\mathbb{A}}$ could be constrained for the $\alpha=3$ case, given a 
gravitational-wave detection consistent with GR. The first column lists the masses of the objects considered, the instrument analyzed and the signal-to-noise ratio (SNR).}
\end{table}
%%%%%%%

Let us now compare these bounds with current constraints. The mass of the graviton has been constrained dynamically to $m_{g} \leq 7.6 \times 10^{-20}$ eV through binary pulsar observations of the orbital period decay and statically to $4.4 \times 10^{-22}$ eV with Solar System constraints (see e.g.~\cite{2011arXiv1107.3528B}). We see then that even with the inclusion of an additional $\mathbb{A}$ parameter, the projected gravitational wave bounds on $m_{g}$ are still interesting. The quantity $\mathbb{A}$ has not been constrained in the gravitational sector. In the electromagnetic sector, the dispersion relation of photon has been constrained: for example, for $\alpha=3$, $\mathbb{A} \lesssim 10^{-25} \; {\rm{eV}}^{-1}$ using TeV $\gamma$-ray observations~\cite{Biller:1998hg}. One should note, however, that such bounds on the photon dispersion relation are independent of those we study here, as in principle the photon and the graviton dispersion relations need not be tied together.
% E/\xi = 4 10^{16} GeV
% A = 2 \xi/Ep

We must stress that this paper deals only with Lorentz-violating corrections to the gravitational wave dispersion relation, and thus, it deals only with {\emph{propagation effects}} and not with {\emph{generation effects}}. Generation effects will in principle be very important, possible leading to the excitation of additional polarizations, as well as modifications to the quadrupole expressions. Such is the case in several modified gravity theories, such as Einstein-Aether theory and Horava-Lifshitz theory~\cite{2011PhRvD..84f4004B, 2004PhRvD..70b4003J, 2007PhRvD..76h4033F, 2010PhRvD..81f4031S, 2009PhRvD..80d4032S, 2011arXiv1105.2555H, 2010AIPC.1241.1128R, 2009PhRvD..79j4004B, 2011arXiv1103.3439M, 2011PhRvD..84j4035P, 2006JCAP...02..003D, 2011arXiv1110.5950Y, 2004PhRvD..70h3509B}. Generically studying the generation problem, however, is difficult as there does not exist a general Lagrangian density that can capture all Lorentz-violating effects. Instead, one would have the gargantuan task of solving the generation problem within each specific theory. 

The goal of this paper, instead, is to consider generic Lorentz-violating effects in the dispersion relation and focus only on the propagation of gravitational waves. This will then allow us to find the corresponding ppE parameters that represent Lorentz-violating propagation. Thus, if future gravitational wave observations peak at these ppE parameters, then one could suspect that some sort of Lorentz-violation could be responsible for such deviations from General Relativistic. Future work will concentrate on the generation problem. 

The remainder of this paper deals with the details of the calculations and is organized as follows. In Sec.~\ref{speedofGW}, we introduce and motivate the modified dispersion relation~(\ref{dispersion}), and derive from it the gravitational-wave speed as a function of energy and the new Lorentz-violating parameters. In Sec.~\ref{PropagationofGW}, we study the propagation of gravitons in a cosmological background as determined by the modified dispersion relation and graviton speed. We find the relation between emission and arrival times of the gravitational waves, which then allows us in Sec.\ref{modifiedWF} to construct a {\emph{restricted}} post-Newtonian (PN) gravitational waveform to $3.5$ PN order in the phase $[{\mathcal O}(v/c)^7]$.  We also discuss the connection to the ppE framework. In Sec.\ref{ConstrainingMG}, we calculate the Fisher information matrix for Ad.~LIGO, ET and a LISA-like mission and determine the accuracy to which the compact binary's parameters can be measured, including a bound on the graviton and Lorentz-violating Compton wavelengths. In Sec.~\ref{conclusions} we present some conclusions and discuss possible avenues for future research.

%%%%%%%%%%%%%%%%%%%%%%%%%%%%%
%%%%%%%%%%%%%%%%%%%%%%%%%%%%%
\section{The Speed of Lorentz-Violating Gravitational Waves}
\label{speedofGW}

In general relativity, gravitational waves travel at the speed of light $c$ because the gauge boson associated with gravity, the graviton, is massless. Modified gravity theories, however, predict modifications to the gravitational-wave dispersion relation, which would in turn force the waves to travel at speeds different than $c$. The most intuitive, yet purely phenomenological modification one might expect is to introduce a mass for the graviton, following the special relativistic relation 
\be
\label{SR}
E^2=p^2c^2+m_{g}^2c^4\,.
\ee
From this dispersion relation, together with the definition $v/c \equiv p/p^0$, or $v \equiv  c^2 p/E$ , one finds the graviton speed~\cite{Will:1997bb} 
\be
\label{SRvelocity}
\frac{v_g^2}{c^2}=1-\frac{m_g^2 c^4}{E^2},
\ee
where $m_g$, $v_{g}$ and $E$ are the graviton's rest mass, velocity and energy. 

Different alternative gravity theories may predict different dispersion relations from Eq.~\eqref{SR}.  A few examples of such relations include the following:
\begin{itemize}
\item {\emph{Double Special Relativity Theory}}~\cite{2001PhLB..510..255A,2002PhRvL..88s0403M,AmelinoCamelia:2002wr,2010arXiv1003.3942A}: $E^2=p^2c^2+m_{g}^2c^4+\eta_{\rm dsrt} E^3 + \ldots$, where $\eta_{\rm dsrt}$ is a parameter of the order of the Planck length. 
\item {\emph{Extra-Dimensional Theories}}~\cite{2011PhLB..696..119S}: $E^2=p^2c^2+m_{g}^2c^4-\alpha_{\rm edt} E^4$, where $\alpha_{\rm edt}$ is a constant related to the square of the Planck length;
\item {\emph{Ho\v{r}ava-Lifshitz Theory}}~\cite{Horava:2008ih,Horava:2009uw,2010arXiv1010.5457V,Blas:2011zd}: $E^{2} = p^{2}c^{2} + (\kappa^{4}_{\rm hl} \mu^{2}_{\rm hl}/16) \; p^{4} + \ldots$, where $\kappa_{\rm hl}$ and $\mu_{\rm hl}$ are constants of the theory; 
\item {\emph{Theories with Non-Commutative Geometries}}~\cite{2011arXiv1102.0117G,Garattini:2011kp,Garattini:2011hy}: $\displaystyle{E^2 g_1^2(E)=m_{g}^2c^4+p^2c^2 g_2^2(E)}$ with $g_2=1$ and $\displaystyle{g_1=(1-\sqrt{\alpha_{\rm ncg}\pi}/2 )\exp({-\alpha_{\rm ncg} {E^2}/{E_p^2})}}$, with $\alpha_{\rm ncg}$ a constant.
\end{itemize} 
Of course, the list above is just representative of a few models, but there are many other examples where the graviton dispersion relation is modified~\cite{Berezhiani:2007zf,Berezhiani:2008nr}. 
In general, a modification of the dispersion relation will be accompanied by a change in either the Lorentz group or its action in real or momentum space. Lorentz-violating effects of this type are commonly found in quantum gravitational theories, including loop quantum gravity~\cite{2008PhRvD..77b3508B} and string theory~\cite{2005hep.th....8119C,2010GReGr..42....1S}.  

Modifications to the standard dispersion relation are usually suppressed by the Planck scale, so one might wonder why one should study them. Recently, Collins, et~al.~\cite{2004PhRvL..93s1301C,2006hep.th....3002C} suggested that Lorentz violations in perturbative quantum field theories could be dramatically enhanced when one regularizes and renormalizes them. This is because terms that would vanish upon renormalization due to Lorentz invariance do not vanish in Lorentz-violating theories, leading to an enhancement after renormalization~\cite{2011arXiv1106.1417G}.

Although this is an appealing argument, we prefer here to adopt a more agnostic viewpoint and simply ask the following question: What type of modifications would enter gravitational-wave observables because of a modified dispersion relation and to what extent can these deviations be observed or constrained by current and future gravitational-wave detectors? In view of this, we postulate the parametrized dispersion relation
of Eq.\ (\ref{dispersion}).

One can see that this model-independent dispersion relation can be easily mapped to all the ones described above, in the limit where $E$ and $p$ are large compared to $m_g$, but small compared to the Planck energy $E_{p}$. More precisely, we have
\begin{itemize}
\item {\emph{Double Special Relativity}}: $ \mathbb{A}  = \eta_{\rm dsrt}$ and $\alpha = 3$. 
\item {\emph{Extra-Dim.~Theories}}: $ \mathbb{A}  = - \alpha_{\rm edt}$ and $\alpha = 4$. 
\item {\emph{Ho\v{r}ava-Lifshitz}}: $ \mathbb{A}  = \kappa^{4}_{\rm hl}  \mu^{2}_{\rm hl}/16$ 
and $\alpha = 4$, but with $m_{g} = 0$.
\item {\emph{Non-Commutative Geometries}}: $ \mathbb{A}  = 2 \alpha_{\rm ncg}/E_{p}^2 $ and $\alpha = 4$, after renormalizing $m_g$ and $c$. 
\end{itemize} 
Of course, for different values of $(\mathbb{A},\alpha)$ we can parameterize other Lorentz-violating corrections to the dispersion relation. One might be naively
tempted to think that a $p^{3}$ or $p^{4}$ correction to the above dispersion relation will induce a $1.5$ or $2$PN correction to the phase relative to the massive graviton term. This, however, would be clearly wrong, as $p$ is the graviton's momentum, not the momentum of the members of a binary system. 

With this modified dispersion relation the modified graviton speed takes the form
\be
\frac{v_{g}^{2}}{c^2}  = 1 - \frac{m_{g}^2 c^4}{E^{2}} - \mathbb{A} E^{\alpha-2} \left (\frac{v}{c} \right )^\alpha\,.
\ee
To first order in $\mathbb{A}$, this can be written as
\be
\frac{v_{g}^{2}}{c^2} = 1 - \frac{m_{g}^2 c^4}{E^{2}} - \mathbb{A} E^{\alpha-2} 
\left(1 - \frac{m_{g}^2 c^4}{E^{2}}\right)^{\alpha/2} \,,
\ee
and in the limit $E \gg m_{g}$ it takes the form
\be
\frac{v_{g}^{2}}{c^2}  = 1 - \frac{m_{g}^2 c^4}{E^{2}} - \mathbb{A} E^{\alpha-2}
\,.
\ee 
Notice that if $\mathbb{A} >0$ or if $m_{g}^{2}c^4/E^{2} > |\mathbb{A}|
E^{\alpha-2}$, then the graviton travels slower than light speed. On
the other hand, if $\mathbb{A} < 0$ and $m_{g}^{2}c^4/E^{2} < |\mathbb{A}|
E^{\alpha-2}$, then the graviton would propagate faster than light
speed.

%%%%%%%%%%%%%%%%%%%%%%%%%%%%%%%%
%%%%%%%%%%%%%%%%%%%%%%%%%%%%%%%%
\section{Propagation of Gravitational Waves with a Modified Dispersion Relation}\label{PropagationofGW}

We now consider the propagation of gravitational waves that satisfy the modified dispersion relation of Eq.~\eqref{dispersion}. Consider  the Friedman-Robertson-Walker background 
\be
ds^2=-dt^2+a^2(t)[d\chi^2+\Sigma^2(\chi)(d\theta^2+\sin^2\theta
\, d\phi^2)],
\ee
where $a(t)$ is the scale factor with units of length, and $\Sigma(\chi)$ is equal to $\chi$, $\sin\chi$ or $\sinh\chi$ if the universe is spatially flat, closed or open, respectively.   Here and henceforth,
we use units with $G=c=1$, where a useful conversion factor is $1 M_\odot=4.925 \times 10^{-6}$ s $= 1.4675$ km. 

In a cosmological background, we will assume that the modified dispersion relation takes the form
\begin{equation}
g_{\mu\nu} p^\mu p^\nu = - m_g^2 - \mathbb{A} |p|^\alpha \,,
\label{dispersionRW}
\end{equation}
where $|p| \equiv (g_{ij} p^i p^j)^{1/2}$.  
Consider a graviton emitted radially at $\chi=\chi_e$ and received at $\chi=0$.  By virtue of the $\chi$ independence of the $t - \chi$ part of the metric, the component $p_\chi$ of its 4-momentum is constant along its worldline. Using $E=p^0$, together with Eq.\ (\ref{dispersionRW}) and the relations
\be
\frac{p^\chi}{E}=\frac{d\chi}{dt},\;\; p^\chi=a^{-2}p_\chi,
\ee
we obtain
\be
\label{speed}
\frac{d\chi}{dt}=-\frac{1}{a}\left [1+\frac{m_g^2 a^2}{p_\chi^2}+\mathbb{A} \left (\frac{ a}{p_\chi}\right)^{2-\alpha} \right ]^{-\frac{1}{2}},
\ee
where $p_\chi^2=a^2(t_e)(E_e^2-m_g^2-\mathbb{A} |p|_e^\alpha)$. The overall minus sign in the above equation is included because the graviton travels from the source to the observer.

Expanding to first order in $(m_g/E_e)\ll 1$, and $\mathbb{A}/p^{2-\alpha} \ll 1$ and integrating from emission time ($\chi=\chi_e$) to arrival time ($\chi = 0$), we find
\ba
\label{distance}
\chi_e&=&\int^{t_a}_{t_e}\frac{dt}{a(t)}-\frac{1}{2}\frac{m_g^2}{a^2(t_e) E^2_e}\int^{t_a}_{t_e}a(t)dt
\nonumber \\
&-&\frac{1}{2} \mathbb{A}\, \biggl(a(t_e) E_e\biggr)^{\alpha-2}\int^{t_a}_{t_e}a(t)^{1-\alpha} dt.
\ea

Consider gravitons emitted at two different times $t_e$ and $t'_e$, with energies $E_e$ and $E'_e$, and received at corresponding arrival times ($\chi_e$ is the same for both). Assuming $\Delta t_e\equiv t_e-t'_e\ll a/\dot{a}$, then
\ba
\label{deltat}
\Delta t_a &=&(1+Z)\left[ \Delta t_e +\frac{D_0}{2\lambda_g^2} \left(\frac{1}{f_e^2}-\frac{1}{{f}_e'{}^2}\right) 
\right.
\nonumber \\
&+& \left.
\frac{D_\alpha}{2 \lambda_{\mathbb{A}}^{2-\alpha}} \; \left(\frac{1}{f_e^{2-\alpha}}-\frac{1}{{f}_e'{}^{2-\alpha}}\right)  \right]\,,
\ea
where $Z\equiv a_0/a(t_e)-1$ is the cosmological redshift, and where we have defined 
\be
\lambda_{\mathbb{A}} \equiv h \; \mathbb{A}^{{1}/{(\alpha-2)}}\,,
\ee
and where $m_g/E_e=(\lambda_g f_e)^{-1}$, with $f_e$ the emitted gravitational-wave frequency, $E_{e} = h f_{e}$ and $\lambda_g=h/m_g\,$ the graviton Compton wavelength. Notice that when $\alpha=2$, then the $\mathbb{A}$ correction vanishes. Notice also that $\lambda_{\mathbb{A}}$ always has units of length, irrespective of the value of $\alpha$.  The distance measure $D_\alpha$ is defined by
\ba
D_\alpha\equiv \left(\frac{1+Z}{a_0}\right)^{1-\alpha}\int_{t_e}^{t_a}a(t)^{1-\alpha} dt
\ea
where $a_0=a(t_a)$ is the present value of the scale factor.
For a dark energy-matter dominated universe $D_\alpha$ and the luminosity distance $D_L$ have the form
\ba
D_\alpha &=& \frac{(1+Z)^{1-\alpha}}{H_0}\int_0^Z\frac{(1+z')^{\alpha-2}dz'}{\sqrt{\Omega_M(1+z')^{3}+\Omega_\Lambda}}\,,
\label{Dalpha-general}
\\
D_L&=&\frac{1+Z}{H_0}\int_0^Z\frac{dz'}{\sqrt{\Omega_M(1+z')^3+\Omega_\Lambda}},
\ea
where  $H_{0} \approx 72 \; {\rm{km}} \; {\rm{s}}^{-1} \; {\rm{Mpc}}^{-1}$ is the value of the Hubble parameter today and $\Omega_M=0.3$ and $\Omega_\Lambda=0.7$ are the matter and dark energy density parameters, respectively. 

Before proceeding, let us comment on the time shift found above in Eq.~\eqref{deltat}. First, notice that this equation agrees with the results of~\cite{Will:1997bb} in the limit $\mathbb{A} \to 0$. Moreover, in the limit $\alpha \to 0$, our results map to those of~\cite{Will:1997bb} with the relation $\lambda^{-2}_{g} \to \lambda_{g}^{-2} + \lambda_{\mathbb{A}}^{-2}$. Second, notice that in the limit $\alpha \to 2$, the $(a(t_{e}) E_{e})^{2-\alpha}$ in Eq.~\eqref{distance} goes to unity and the $\mathbb{A}$ correction becomes frequency independent. This makes sense, since in that case the Lorentz-violating correction we have introduced acts as a renormalization factor for the speed of light. 

%%%%%%%%%%%%%%%%%%%%%%%%%%%%%%%%
%%%%%%%%%%%%%%%%%%%%%%%%%%%%%%%%
\section{Modified Waveform in the Stationary Phase Approximation}\label{modifiedWF}

We consider the gravitational-wave signal generated by a non-spinning, quasi-circular inspiral in the post-Newtonian approximation. In this scheme, one assumes that orbital velocities are small compared to the speed of light ($v \ll 1$) and gravity is weak ($m/r \ll1$).  Neglecting any amplitude corrections (in the so-called {\emph{restricted}} PN approximation), the plus- and cross-polarizations of the metric perturbation can be represented as
\ba
\label{eq:h(t)}
h(t)&\equiv&A(t) e^{-i\Phi(t)},\\
\Phi(t)&\equiv&\Phi_c+2\pi\int^t_{t_c}f(t)dt,
\ea
where $A(t)$ is an amplitude that depends on the gravitational-wave polarization (see e.g.~Eq. $(3.2)$ in~\cite{Will:1997bb}), while $f(t)$ is the observed gravitational-wave frequency, and $\Phi_c$ and $t_c$ are a {\emph{fiducial}} phase and fiducial time, respectively, sometimes called the coalescence phase and time. 

The Fourier transform of Eq.~\eqref{eq:h(t)} can be obtained analytically in the stationary-phase approximation, where we assume that the phase is changing much more rapidly than the amplitude~\cite{Droz:1999qx,Yunes:2009yz}.  We then find
\be
\tilde{h}({f})=\frac{\tilde{A}({t})}{\sqrt{\dot{f}({t})}} e^{i\Psi({f})} \,,
\ee
where $f$ is the gravitational-wave frequency at the detector and
\ba
\tilde{A}({t})&=&\frac{4}{5}\frac{\mathcal{M}_e}{a_0 \Sigma(\kappa_e)} (\pi \mathcal{M}_e {f}_e)^{2/3},\\\label{phase}
\Psi({f})&=&2\pi\int_{f_c}^{{f}}(t-t_c)df+2\pi{f}t_c-\Phi_c-\frac{\pi}{4}.
\ea
In these equations, $\mathcal{M}_e=\eta^{3/5} m$ is the {\emph{chirp}} mass of the source, where $\eta = m_1m_2/(m_1+m_2)$ is the symmetric mass ratio. 

We can now substitute Eq.~(\ref{deltat}) into Eq.~(\ref{phase}) to relate the time at the detector to that at the emitter. Assuming that $\alpha \neq 1$, we find 
\ba
\Psi_{\alpha \neq 1}({f})&=& 2\pi \int_{f_{ec}}^{{f_e}}(t_e-t_{ec})df_e  -\frac{\pi D_0}{f_e \lambda_g^2}   
\\ \nonumber 
&-&
\frac{1}{(1-\alpha)} \frac{\pi D_{\alpha}}{f_{e}^{1-\alpha}\lambda_{\mathbb{A}}^{2-\alpha} } 
+2\pi{f} \bar{t}_c-\bar{\Phi}_c-\frac{\pi}{4},
\ea
while for  $\alpha=1$, we find 
\ba
\label{alpha1-dephasing}
\Psi_{\alpha=1}({f})&=& 2\pi \int_{f_{ec}}^{{f_e}}(t_e-t_{ec})df_e  -\frac{\pi D_0}{f_e \lambda_g^2}   
 \\ \nonumber
&+&
\frac{\pi D_{1}}{\lambda_{\mathbb{A}}} \ln\left(\frac{f_{e}}{f_{ec}}\right) 
+2\pi{f} \bar{\bar{t}}_c - \bar{\bar{\Phi}}_c-\frac{\pi}{4}\,.
\ea
The quantities $(\bar{t}_{c},\bar{\bar{t}}_{c})$ and $(\bar{\phi}_{c},\bar{\bar{\phi}}_{c})$ are new coalescence times and phases, into which constants of integration have been absorbed.  

We can relate $t_e - t_{ec}$ to $f_e$ by integrating the frequency chirp equation for non-spinning, quasi-circular inspirals from general relativity~\cite{Will:1997bb}:
\ba
\frac{df_e}{dt_e}&=&\frac{96}{5\pi\mathcal{M}_e^2} (\pi \mathcal{M}_e f_e)^{11/3}\biggl[ 1-\biggl(\frac{743}{336} +\frac{11}{4} \eta \biggr) (\pi m f_e)^{2/3} 
\nonumber \\
&+& 
4\pi (\pi m f_e)\biggr],
\ea
where we have kept terms up to $1$PN order.  In the calculations that follow, we actually account for corrections up to $3.5$PN order, although we don't show these higher-order terms here (they can be found e.g.~in~\cite{2009PhRvD..80h4043B}). 

After absorbing further constants of integration into $(\bar{t}_c,\bar{\Phi}_c,\bar{\bar{t}}_c,\bar{\bar{\Phi}}_c)$, dropping the bars, and re-expressing everything in terms of the \emph{measured} frequency ${f}$ at the detector [note that $\dot{f}^{1/2}=(df_e/dt_e)^{1/2}/(1+Z)$], we obtain 
\ba\label{waveform}
\tilde{h}({f})=\left\{ \begin{array}{cc}
\tilde{A}({f}) e^{i\Psi({f})},& \mbox{for $0<{f}<{f}_{max}$}\\
0,&\mbox{for ${f}>{f}_{max}$} \,,
\end{array}
\right.
\ea
with the definitions
\begin{align}
\tilde{A}({f}) &\equiv \epsilon \; \mathcal{A} \; {u}^{-7/6}\,,
\qquad
{\mathcal{A}} =\sqrt{\frac{\pi}{30}}\frac{\mathcal{M}^2}{D_L}\,,
\\ \nonumber 
\label{psi}
\Psi({f})&= \Psi_{\GR}({f}) + \delta \Psi({f})\,,
\\
\Psi_{\GR}(f) &=2\pi{f}t_c-\Phi_c -\frac{\pi}{4}+\frac{3}{128}u^{-5/3}  
\nonumber \\ 
&\times \sum_{n=0}^{\infty} \left[c_{n} + \ell_{n} \ln(u) \right] u^{n/3}\,,
\end{align}
where the numerical coefficient $\epsilon = 1$ for LIGO and ET, but $\epsilon = \sqrt{3}/2$ for a LISA-like mission (because when one angle-averages, the resulting geometric factors depend slightly on the geometry of the detector). The coefficients $(c_{n},\ell_{n})$ can be read up to $n=7$ for example from Eq.~$(3.18)$ in~\cite{2009PhRvD..80h4043B}. In these equations, $u\equiv\pi\mathcal{M}{f}$ is a dimensionless frequency, while $\mathcal{M}$ is the measured chirp mass, related to the source chirp mass by $\mathcal{M}=(1+Z)\mathcal{M}_e$. The frequency ${f}_{max}$ represents an upper cut-off frequency where the PN approximation fails.

The dephasing caused by the propagation effects takes a slightly different form depending on whether  $\alpha \neq 1$  or  $\alpha=1$. In the general $\alpha \neq 1$ case, we find 
\be
\delta \Psi_{\alpha \neq 1}(f) = -\beta u^{-1} - \zeta u^{\alpha-1}\,,
\label{Dephasing-full}
\ee
where the parameters $\beta$ and $\zeta$ are given by
\ba\label{beta}
\beta&\equiv&\frac{\pi^2 D_0\mathcal{M}}{\lambda_g^2(1+Z)},
\\ \label{zeta}
\zeta_{\alpha \neq 1}&\equiv& \frac{\pi^{2-\alpha}}{(1-\alpha)} \frac{D_\alpha}{\lambda_{\mathbb{A}}^{2-\alpha}}  \frac{\mathcal{M}^{1-\alpha}}{(1+Z)^{1-\alpha}}\,. 
\\\nonumber&&
\ea
In the special $\alpha=1$ case, we find
\be
\delta \Psi_{\alpha = 1}(f) = -\beta u^{-1}  
+ \zeta_{\alpha=1} \ln\left(u\right)\,,
\label{newpsi}
\ee
where $\beta$ remains the same, while 
\be
\zeta_{\alpha=1} = \frac{\pi D_{1}}{\lambda_{\mathbb{A}}}\,,
\ee
and we have re-absorbed a factor into the phase of coalescence. 

As before, notice that in the limit $\mathbb{A} \to 0$, Eq.~\eqref{Dephasing-full} reduces to the results of~\cite{Will:1997bb} for a massive graviton. Also note that, as before, in the limit $\alpha \to 0$, we can map our results to those of~\cite{Will:1997bb} with $\lambda_{g}^{-2} \to \lambda_{g}^{-2} + \lambda_{\mathbb{A}}^{-2}$, i.e.~in this limit, the mass of the graviton and the Lorentz-violating $\mathbb{A}$ term become $100\%$ degenerate. In the limit $\alpha \to 2$, Eq.~\eqref{deltat} becomes frequency-independent, which then implies that its integral, Eq.~\eqref{phase}, becomes linear in frequency, which is consistent with the $\alpha \to 2$ limit of Eq.~\eqref{Dephasing-full}. Such a linear term in the gravitational-wave phase can be reabsorbed through a redefinition of the time of coalescence, and thus is not observable. This is consistent with the observation that the dispersion relation with $\alpha=2$ is equivalent to the standard massive graviton one with a renormalization of the speed of light. When $\alpha=1$, Eq.~\eqref{deltat} leads to a $1/f$ term, whose integral in Eq.~\eqref{phase} leads to a $\ln(f)$ term, as shown in Eq.~\eqref{alpha1-dephasing}. Finally, notice that, in comparision with the phasing terms that arise in the PN approximation to standard general relativity, these corrections are effectively of $(1+3\alpha/2)$PN order, which implies that the $\alpha=0$ term leads to a 1PN correction as in~\cite{Will:1997bb}, the $\alpha=1$ case leads to a $2.5$PN correction, the $\alpha=3$ case leads to a $5.5$PN correction and $\alpha=4$ leads to a $7$PN correction. This suggests that the accuracy to constrain $\lambda_{\mathbb{A}}$ will deteriorate very rapidly as $\alpha$ increases.

%----------------------------------------------------------------------------
\subsection*{Connection to the Post-Einsteinian Framework}

Recently, there has been an effort to develop a framework suitable for testing for deviations from general relativity in gravitational-wave data.  In analogy with the parametrized post-Newtonian (ppN) framework~\cite{1971ApJ...163..611W,1972ApJ...177..775N,1972ApJ...177..757W,1973ApJ...185...31W,lrr-2006-3,Will:1993ns}, the parametrized post-Einsteinian (ppE) framework~\cite{Yunes:2009ke,Vigeland:2011ji,Cornish:2011ys} suggests that we deform the gravitational-wave observable away from our GR expectations in a well-motivated, parametrized fashion. In terms of the Fourier transform of the waveform observable in the SPA, the simplest ppE meta-waveform is
\be
\tilde{h}_{\ppE}(f) = \tilde{A}_{\GR} \left(1+\alpha_{\ppE} u^{a_{\ppE}} \right) e^{i \Psi_{\GR}(f) + i \beta_{\ppE}\;u^{b_{\ppE}}}\,,
\ee
where $(\alpha_{\ppE},a_{\ppE},\beta_{\ppE},b_{\ppE})$ are ppE, theory parameters. Notice that in the limit $\alpha_{\ppE} \to 0$ or $\beta_{\ppE} \to 0$, the ppE waveform reduces exactly to the SPA GR waveform. The proposal is then to match-filter with template families of this type and allow the data to select the best-fit ppE parameters to determine whether they are consistent with GR. 

We can now map the ppE parameters to those obtained from a generalized, Lorentz-violating dispersion relation:
\ba
\alpha_{\ppE} &=& 0\,
\qquad
\beta_{\ppE} = - \zeta
\qquad
b_{\ppE} = \alpha - 1\,.
\ea
Quantum-gravity inspired Lorentz-violating theories suggest modified dispersion exponents $\alpha=3$ or $4$, to leading order in $E/m_{g}$, which then implies ppE parameters $b_{\ppE} = 2$ and $3$. Therefore, if after a gravitational wave has been detected, a Bayesian analysis with ppE templates is performed that leads to values of $b_{\ppE}$ that peak around $2$ or $3$, this would indicate the possible presence of Lorentz violation~\cite{Cornish:2011ys}. Notice however that the $\alpha=1$ case cannot be recovered by the ppE formalism without generalizing it to include $\ln{u}$ terms. Such effects are analogous to memory corrections in PN theory.

At this point, we must spell out an important caveat.  The values of $\alpha$ that represent Lorentz violation for quantum-inspired theories ($\alpha=3,4$) correspond to very high PN order effects, i.e.~a relative $5.5$ or $7$ PN correction respectively. Any gravitational-wave test of Lorentz violation that wishes to constrain such steep momentum dependence would require a very accurate (high PN order) modeling of the general relativistic waveform itself. In the next section, we will employ $3.5$ PN accurate waveforms, which are the highest-order known, and then ask how well $\zeta$ and $\beta$ can be constrained. Since we are neglecting higher than $3.5$ PN order terms in the template waveforms, we are neglecting also any possible correlations or degeneracies between these terms and the Lorentz-violating terms. Therefore, any estimates made in the next section are at best optimistic bounds on how well gravitational-wave measurements could constrain Lorentz violation.

%%%%%%%%%%%%%%%%%%%%%%%%%%%%%
%%%%%%%%%%%%%%%%%%%%%%%%%%%%%
\section{Constraining a Modified Graviton Dispersion Relation}\label{ConstrainingMG}

In this section, we perform a simplified Fisher analysis, following the 
method outlined for compact binary inspiral in~\cite{cutflan94,finn,Poisson:1995ef},
to get a sense of the bounds one could place on $(\lambda_{g},\lambda_{\mathbb{A}})$
given a gravitational-wave detection that is consistent with general relativity. We begin by summarizing some of 
the basic ideas behind a Fisher analysis, introducing some notation. We then apply 
this analysis to an Adv.~LIGO detector, an ET detector and a LISA-like mission. 

%------------------------------------------------
\subsection{General Considerations}

Given a noise power spectrum, $S_n(f)$, we can define the inner product of
signals $h_1$ and $h_2$ as 
\begin{equation}
(h_1|h_2) \equiv 2\int_0^\infty {{\tilde h_1^* \tilde h_2 +\tilde
h_2^* \tilde h_1 } \over S_n(f)} df \,,
\label{innerproduct}
\end{equation}
where $\tilde h_1$ and $\tilde h_2$ are the Fourier transforms of signals $1$ and $2$ respectively and star superscript stands for complex conjugation.
The SNR for a given signal $h$ is simply
\begin{equation}
\rho[h] = (h|h)^{1/2} \,.
\label{signaltonoise}
\end{equation}
If the signal depends on a set of parameters $\theta^a$ that we wish to 
estimate via matched filtering, then the root-mean-square error on parameter $\theta^a$
in the limit of large SNR is (no summation over $a$ implied here)
\begin{equation}
\Delta \theta^a \equiv \sqrt{\langle (\theta^a - \langle \theta^a
\rangle )^2 \rangle } = \sqrt{ \Sigma^{aa}} \,.
\label{deltatheta}
\end{equation}
The quantity $\Sigma^{aa}$ is the $(a,a)$ component of the variance-covariance matrix,  
which is the inverse of the Fisher information matrix, $\Gamma_{ab}$, defined as  
\begin{equation}
\Gamma_{ab} \equiv \left ( {{\partial h} \over {\partial \theta^a}}
\big | {{\partial h} \over {\partial \theta^b}} \right ) \,.
\label{fisher}
\end{equation}
The off-diagonal elements of the variance-covariance matrix give the parameter correlation coefficients, which we define as 
\begin{equation}
c^{ab} \equiv \Sigma^{ab}/\sqrt{\Sigma^{aa}\Sigma^{bb}} \,.
\label{correlation} 
\end{equation}

We will work with an angle-averaged response function, so that the templates depend only on the parameters: 
\be
\theta^{a}= (\ln {\cal A}, \Phi_c, f_0
t_c, \ln {\cal M}, \ln \eta, \beta,\zeta)\,,
\ee 
where each component of the vector $\theta^{a}$ is dimensionless.
We recall that ${\cal{A}}$ is an overall amplitude that contains information about the gravitational-wave polarization and the beam-pattern function angles. The quantities $\Phi_{c}$ and $t_{c}$ are the phase and time of coalescence, where $f_0$ is a frequency
characteristic of the detector, typically a ``knee'' frequency, or a
frequency at which $S_n(f)$ is a minimum. The parameters ${\cal{M}}$ and $\eta$ are the chirp mass and symmetric mass ratio, which characterize the compact binary system under consideration. The parameters $(\beta,\zeta)$ describe the  massive graviton and Lorentz-violating terms respectively.
   
The SNR for the templates in Eq.~\eqref{waveform} is simply 
\be
\rho= 2 \; \epsilon \; \mathcal{A} \; \left({\cal{M}} \pi \right)^{-7/6} \; f_0^{-2/3} \; I(7)^{1/2} S_0^{-1/2}\,,
\label{SNR}
\ee
where we have defined the integrals 
\be
\label{moment}
I(q)\equiv \int_0^\infty \frac{x^{-q/3}}{g(x)},
\ee
with $x \equiv f/f_{0}$. The quantity $g(x)$ is the rescaled power spectral density, defined via $g (x) \equiv S_h(f)/S_{0}$ for the detector in question, and $S_{0}$ is an overall constant. When computing the Fisher matrix, we will replace the amplitude ${\cal{A}}$ in favor of the SNR, using Eq.~\eqref{SNR}. This will then lead to bounds on $(\beta,\zeta)$ that depend on the SNR and on a rescaled version of the moments $J(q)\equiv I(q)/I(7)$.

In the next subsections, we will carry out the integrals in Eq.~\eqref{moment}, but we will approximate the limits of integration by certain $x_{\rm min}$ and $x_{\rm max}$~\cite{Berti:2004bd}. The maximum frequency will be chosen to be the smaller of a certain instrumental maximum threshold frequency and that associated with a gravitational wave emitted by a particle in an innermost-stable circular orbit (ISCO) around a Schwarzschild black hole (BH): $f_{\rm max} = 6^{-3/2} \pi^{-1} \eta^{3/5} {\cal{M}}^{-1}$. The maximum instrumental frequency will be chosen to be $(10^{5},10^{3},1)$ Hz for Ad.~LIGO, ET and LISA-like, respectively. The minimum frequency will be chosen to be the larger of a certain instrumental minimum threshold frequency and, in the case of a space mission, the frequency associated with a gravitational wave emitted by a test-particle one year prior to reaching the ISCO. The minimum instrumental frequency will be chosen to be $(10,1,10^{-5})$ Hz for Ad.~LIGO, ET and a LISA-like mission, respectively.

Once the Fisher matrix has been calculated, we will invert it using a Cholesky decomposition to find the variance-covariance matrix, the diagonal components of which give us a measure of the accuracy to which parameters could be constrained. Let us then define the upper bound we could place on $(\beta,\zeta)$ as $\Delta \beta \equiv \Delta^{1/2} /\rho$ and $\Delta \zeta \equiv \bar\Delta^{1/2} /\rho$, where $\Delta$ and $\bar\Delta$ are numbers. Combining these definitions with Eqs. (\ref{beta}) and (\ref{zeta}), we find, for $\alpha \ne 1$, the
bounds:
\ba
\lambda_g &>& \sqrt{ \frac{\rho \, D_0 \, {\cal M}}{(1+Z)}} \frac{\pi}{\Delta^{1/4}} \,,
\label{lambdabound}
\\
\lambda_{\mathbb{A}}^{\alpha-2} &<&\frac{|1-\alpha|}{\pi^{2-\alpha}} \frac{\bar{\Delta}^{1/2}}{D_{\alpha} \rho} 
\frac{{\cal{M}}^{\alpha-1}}{(1 + Z)^{\alpha-1}}  \,,
\ea
Notice that the direction of the bound on $\lambda_{\mathbb{A}}$ itself depends on whether $\alpha > 2$ or $\alpha < 2$; but because $\mathbb{A} = (\lambda_{\mathbb{A}}/h)^{\alpha-2}$,  all cases yield an upper bound on $\mathbb{A}$. For the case $\alpha=1$
, we find
\ba
\lambda_{\mathbb{A}_{\alpha=1}} &>& \frac{\pi D_{1}}{\bar{\Delta}^{1/2}} \rho\,,
\ea
In the remaining subsections, we set $\beta=0$ and $\zeta=0$ in all partial derivatives when computing the Fisher matrix, since we derive the error in estimating  $\beta$ and $\zeta$ about the nominal or {\it a priori} general relativity values, $(\beta,\zeta)=(0,0)$. 

%%%%%%%%%%%%%%%
%%%% RESULTS %%%%%
%%%%%%%%%%%%%%%

%------------------------------------------------
\subsection{Detector Spectral Noise Densities}

\begin{figure*}[ht!]
\begin{tabular}{cc}
\includegraphics[width=9cm,clip=true]{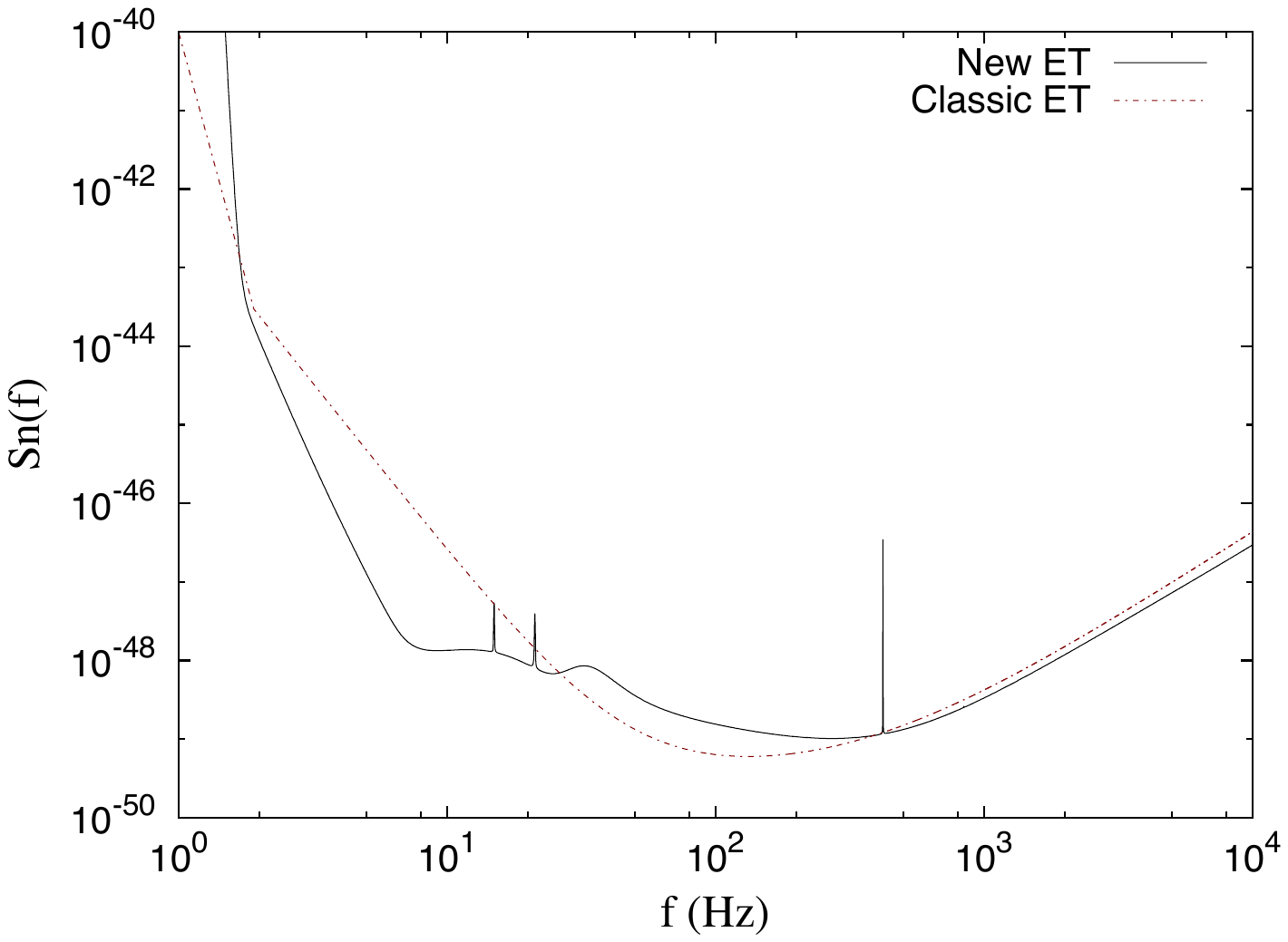} &
\includegraphics[width=9cm,clip=true]{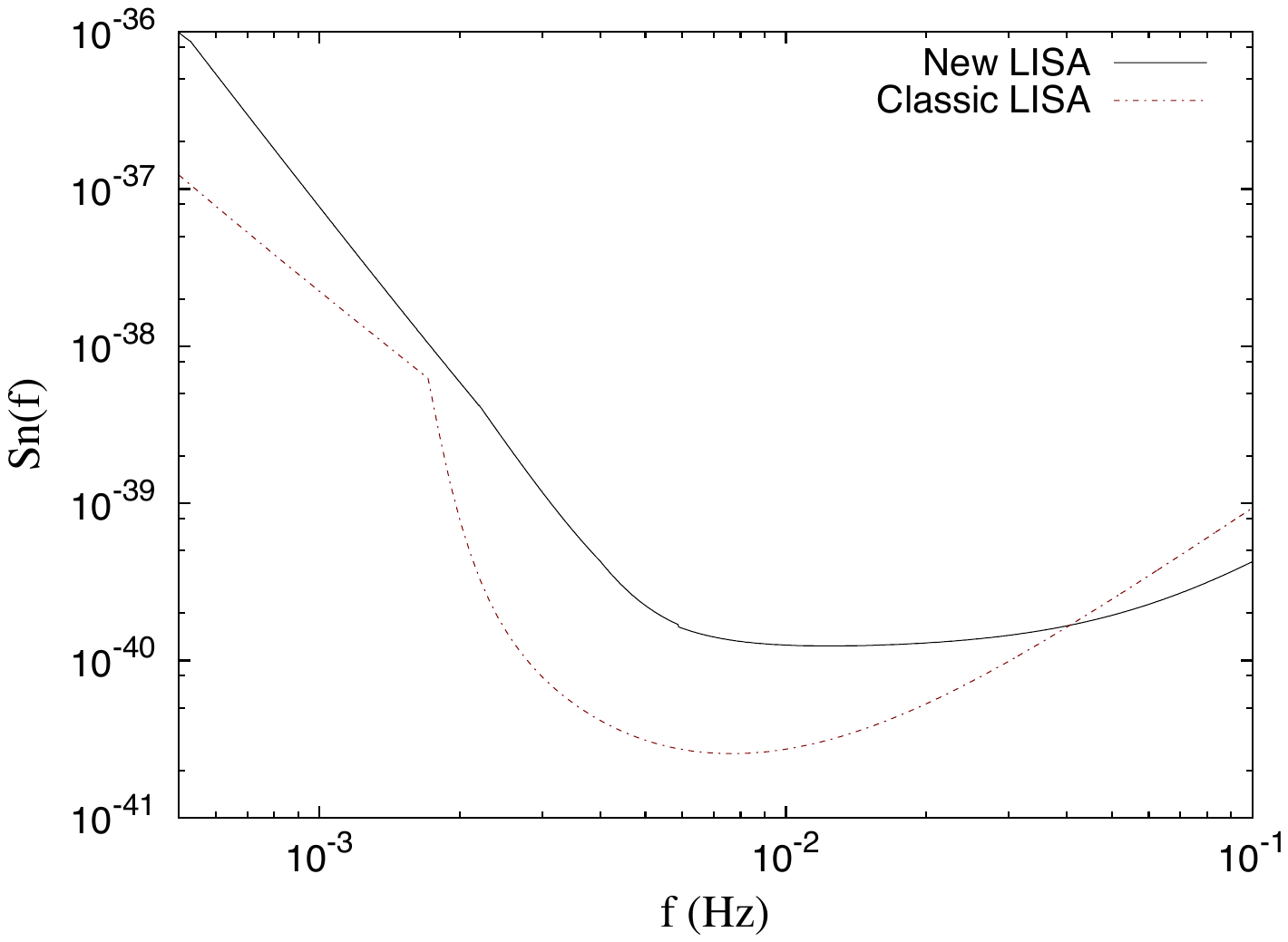}
\end{tabular}
\caption{\label{Noisecurves} ET (left) and LISA (right) spectral noise density curves for the classic design (dotted) and the new NGO design (solid).}
\end{figure*}

We model the Ad.~LIGO spectral noise density via~\cite{Mishra:2010tp}
\ba
\frac{S_h(f)}{S_{0}} = \left\{ \begin{array}{cc}
10^ {16-4(x f_0-7.9)^2}+2.4\times 10^{-62} x^{-50} 
\\ 
+0.08 x^{-4.69}
\\  
+123.35\biggl(\frac{1-0.23 x^2+0.0764 x^4}{1+0.17 x^2}\biggr), & f \geq f_s, \\
\infty, & f < f_s,
\end{array}
\right.
\label{advligo}
\ea
Here, $f_0=215$ Hz, $S_0=10^{-49}$ Hz$^{-1}$, and $f_s=10\; {\rm{Hz}}$ is a low-frequency cutoff below which $S_h(f)$ can be considered infinite for all practical purposes

The initial ET design postulated the spectral noise density~\cite{Mishra:2010tp}
\ba
\frac{S_h(f)}{S_{0}}  =  \left\{ \begin{array}{cc}
\left[a_1 x^{b_1}+a_2 x^{b_2} +a_3 x^{b_3}+a_4 x^{b_4}\right]^{2}, & f \geq  f_s\, \\
\infty, & f < f_s,
\end{array}
\right.
\label{et}
\ea
where $f_0=100\,\mbox{Hz}$, $S_0 = 10^{-50}\,\mbox{Hz}^{-1}$, $f_s=1 \; {\rm{Hz}}$, and 
\begin{eqnarray}
a_1 =&2.39\times10^{-27},~\quad
b_1 &= -15.64,\nonumber\\
a_2 &= 0.349, \quad\quad\quad\quad
b_2 &= -2.145,\nonumber\\
a_3 &= 1.76, ~~\quad\quad\quad\quad
b_3 &= -0.12,\nonumber\\
a_4 &= 0.409, \quad\quad\quad\quad
b_4 &= 1.10.
\label{constants_sqrtpsd}
\end{eqnarray}

The classic LISA design had an approximate spectral noise density curve that could be
modeled via (see eg.~\cite{Berti:2004bd, Barack:2003fp}):
\ba\nonumber\label{lisa}
S_h(f)&=&{\rm min}\biggl\{
S_h^{\rm NSA}(f)/{\rm exp}
\left(-\kappa T^{-1}_{\rm mission} dN/df\right),\\
&&~
S_h^{\rm NSA}(f)+S_h^{\rm gal}(f)
\biggr\}+S_h^{\rm ex-gal}(f)\,.
\label{Shtot}
\ea
where 
\ba\nonumber
S_h^{\rm NSA}(f)&=& \biggl[9.18\times 10^{-52}\left(\f{f}{1~{\rm Hz}}\right)^{-4}
+1.59\times 10^{-41}\\
&&+9.18\times 10^{-38}\left(\f{f}{1~{\rm Hz}}\right)^2\biggr]~{\rm Hz}^{-1}\,.
\\
S_h^{\rm gal}(f) &=&
2.1\times 10^{-45}\left(\f{f}{1~{\rm Hz}}\right)^{-7/3}~{\rm Hz}^{-1}\,,
\\
S_h^{\rm ex-gal}(f) &=&
4.2\times 10^{-47}\left(\f{f}{1~{\rm Hz}}\right)^{-7/3}~{\rm Hz}^{-1}\,.
\ea
and 
\be
\f{dN}{df}=2\times 10^{-3}~{\rm Hz}^{-1}
\left(\f{1~{\rm Hz}}{f}\right)^{11/3}\,;
\ee
with $\Delta f=T^{-1}_{\rm mission}$ the bin size of the discretely
Fourier transformed data for a classic LISA mission lasting a time
$T_{\rm mission}$ and $\kappa\simeq 4.5$ the average number of
frequency bins that are lost when each galactic binary is fitted
out. 

Recently, the designs of LISA and ET have changed somewhat. The new spectral noise density curves can be computed numerically~\cite{2011CQGra..28i4013H,Sathya,Berti} and are plotted in Fig.~\ref{Noisecurves}. Notice that the bucket of the NGO noise curve has shifted to higher frequency, while the new ET noise curve is more optimistic than the classic one at lower frequencies. The spikes in the latter are due to physical resonances, but these will not affect the analysis. In the remainder of this paper, we will use the new ET and NGO noise curves to estimate parameters.  

%------------------------------------------------
\subsection{Results}

\begin{figure*}[htb]
\begin{tabular}{rcl}
\hspace{-0.5cm} \includegraphics[width=6.25cm,clip=true]{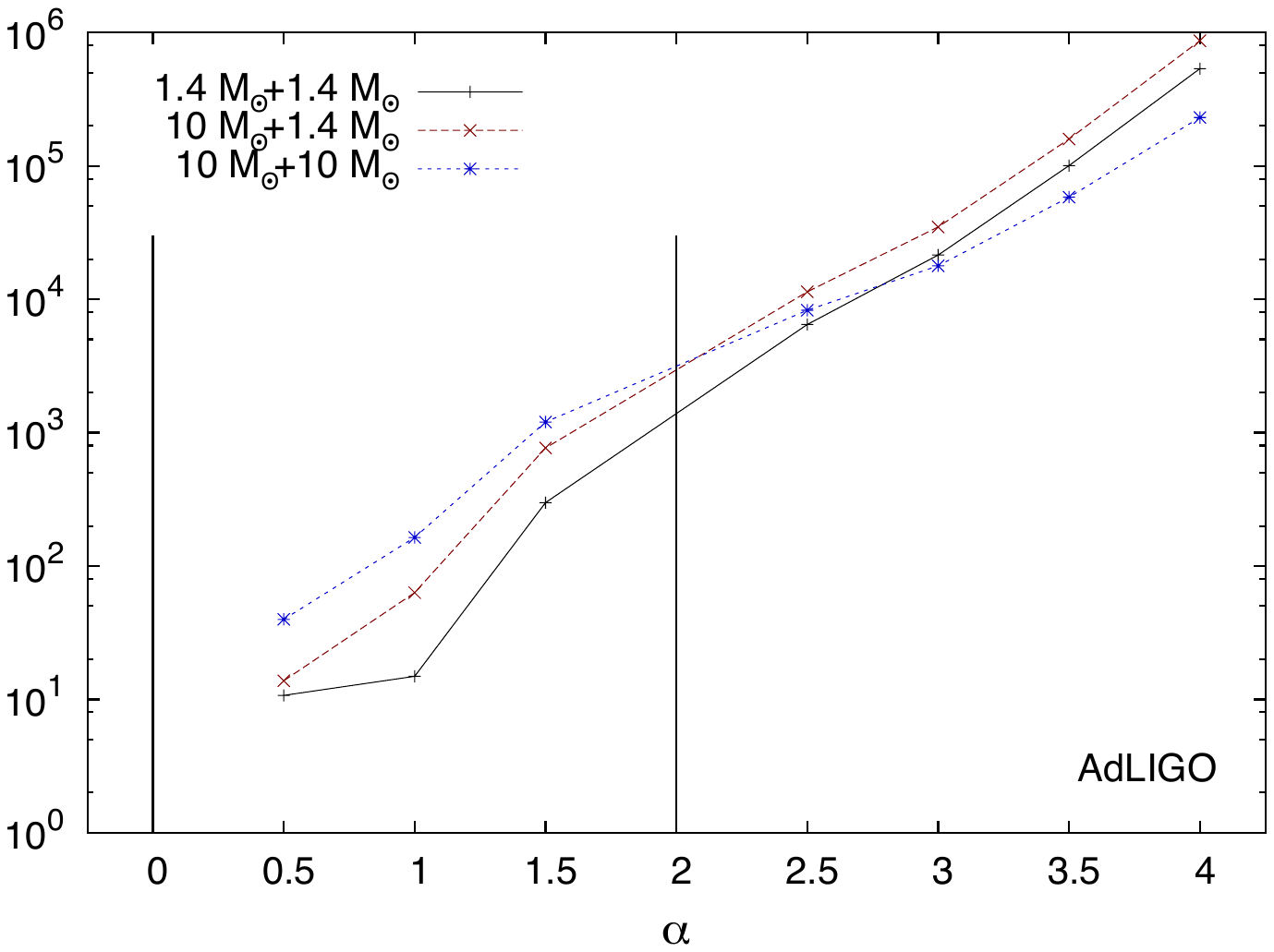} &
\hspace{-0.5cm} \includegraphics[width=6.25cm,clip=true]{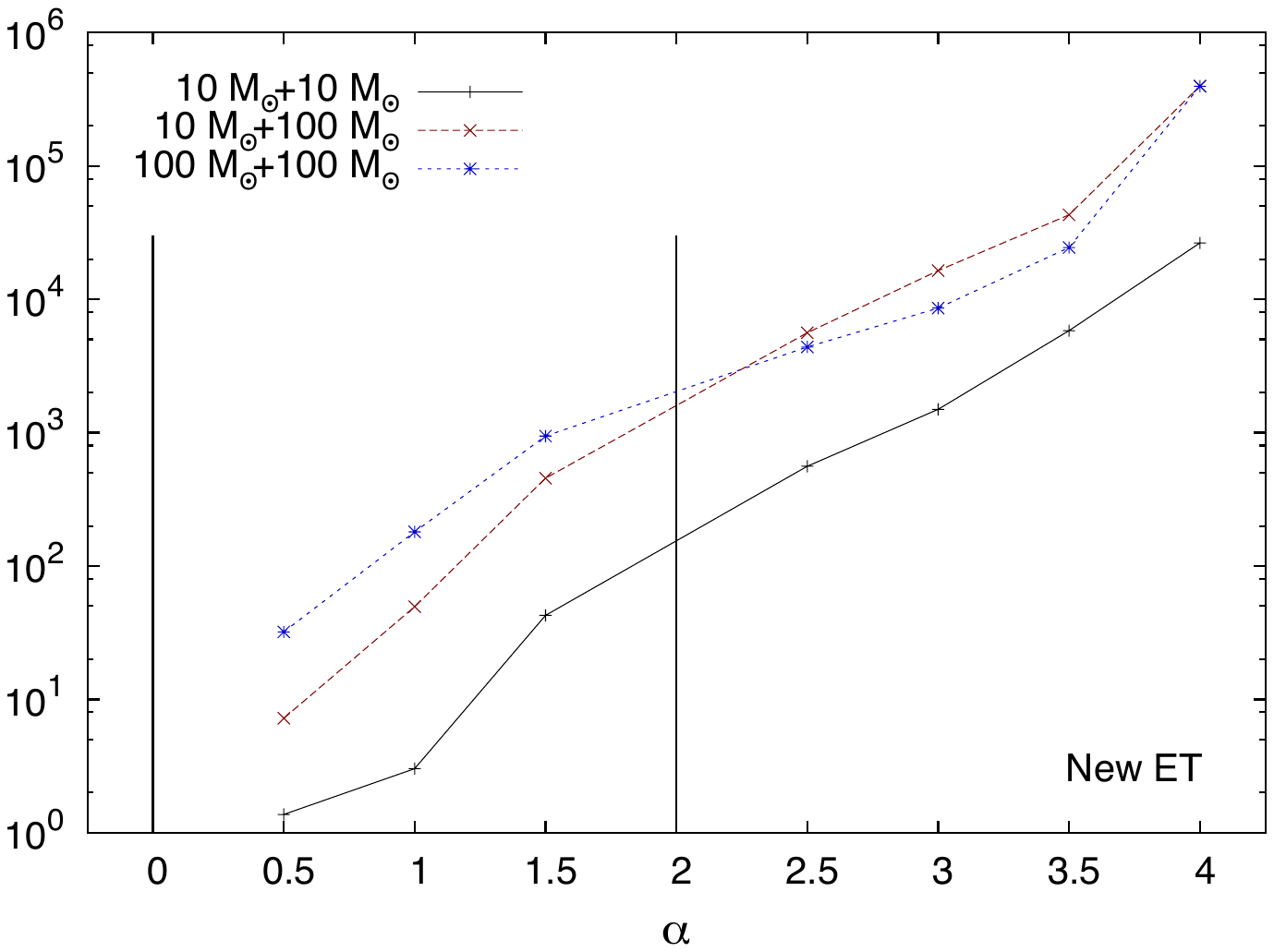} &
\hspace{-0.5cm} \includegraphics[width=6.25cm,clip=true]{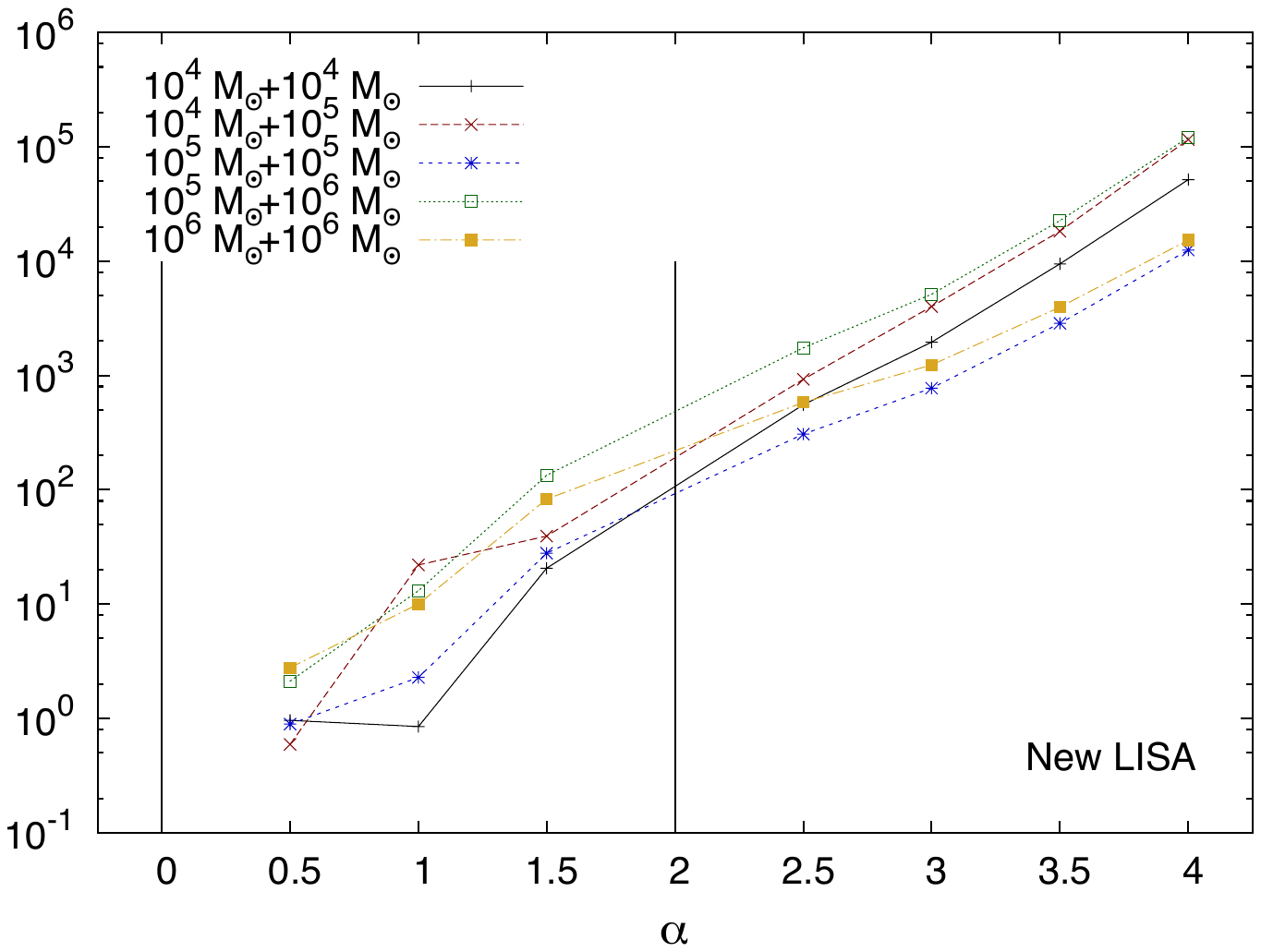}
\end{tabular}
\caption{\label{Zeta-figs} Bounds on parameter $\zeta$ for different values of $\alpha$, using AdLIGO and $\rho=10$ (left panel), ET and $\rho=50$ (center panel), and NGO and $\rho = 100$ (right panel). Vertical lines at $\alpha=(0,2)$ show where the $\zeta$ correction becomes $100\%$ degenerate with other parameters. Each panel contains several curves that show the bound for systems with different masses.}
\end{figure*}
We plot the bounds that can be  placed on $\zeta$ in Fig.~\ref{Zeta-figs} as a function of the $\alpha$ parameter. The left panel corresponds to the bounds placed with Ad.~LIGO and $\rho = 10$ ($D_{L} \sim 160 \; {\rm{Mpc}}$, $Z \sim 0.036$ for a double neutron-star inspiral), the middle panel corresponds to ET and $\rho = 50$ ($D_{L} \sim 2000 \; {\rm{Mpc}}$, $Z \sim 0.39$ for a double $10 M_{\odot}$ BH inspiral) and the right panel corresponds to NGO and $\rho = 100$ ($D_{L} \sim 20,000 \; {\rm{Mpc}}$, $Z \sim 2.5$ for a double $10^{5} M_{\odot}$ BH inspiral).  When $\alpha=0$ or $\alpha=2$, $\zeta$ cannot be measured at all, as it becomes $100\%$ correlated with either standard massive graviton parameters.  Thus we have drawn vertical lines in those cases.  As the figure clearly shows, the accuracy to which $\zeta$ can be measured deteriorates rapidly as $\alpha$ becomes larger. In fact, once $\alpha > 4$, we find that $\zeta$ cannot be confidently constrained anymore because the Fisher matrix becomes non-invertible (it's condition number exceeds $10^{16}$).

Attempting to constrain values of $\alpha > 5/3$ becomes problematic not just from a data analysis point of view, but also from a fundamental one. The PN templates that we have constructed contain general relativity phase terms up to $3.5$ PN order. Such terms scale as $u^{2/3}$, which corresponds to $\alpha = 5/3$. Therefore, trying to measure values of $\alpha \geq 5/3$ without including the corresponding 4PN and higher-PN order terms is not well-justified. We have done so here, neglecting any correlations between these higher order PN terms and the Lorentz-violating terms, in order to get a rough sense of how well Lorentz-violating modifications could be constrained.

%%%%%%%
\begin{table}[ht]
\hsize\textwidth\columnwidth\hsize\csname
           @twocolumnfalse\endcsname
\begin{tabular}{c c c || c c c c c c c c c c c c}
\hline\hline
${\rm{Detector}}$ &$m_1$&$m_2$&$\Delta \phi_c$&$\Delta t_c$&$\Delta {\cal M}/{\cal M}$
&$\Delta \eta/\eta$&$\Delta\lambda_g$&$\Delta\lambda_\mathbb{A}$&$c_{{\cal M}\eta}$&$c_{{\cal M}
\beta}$&$c_{\eta \beta}$&$c_{{\cal M}\zeta}$&$c_{\eta
\zeta}$&$c_{\beta \zeta}$\\
\hline
{\rm{Ad.~LIGO}} & 1.4&1.4&3.61&1.80&0.0374\%&6.80\%&3.34&0.911&-0.962&-0.991&0.989&-0.685&0.803&0.740\\
&1.4&10&3.34&9.99&0.267\%&12.8\%&3.48&4.36&-0.977&-0.993&0.917&-0.830&0.923&0.875\\
& 10&10&4.16&31.0&2.40\%&72.2\%&3.53&8.40&-0.978&-0.994&0.995&-0.874&0.947&0.915\\
\hline\hline
{\rm{ET}} & 10&10&       0.528&1.59&0.0174\%&1.70\%&4.15&0.0286&-0.952&-0.986&0.988&-0.742&0.875&0.813\\
& 10&100&    1.12&44.5&0.259\%&6.67\%&2.58&1.38&-0.974&-0.993&0.993&-0.872&0.951&0.915\\
& 100&100&  5.23&203&4.03\%&67.6\%&1.86&4.12&-0.983&-0.995&0.996&-0.914&0.969&0.947\\
\hline\hline
{\rm{NGO}} 
& $10^4$&$10^4$&  0.264&1.05&0.00124\%&0.368\% &4.06& 0.266 	&-0.957&-0.990&0.986&-0.636&0.761&0.687\\
& $10^4$&$10^5$&  0.264&5.42&0.00434\%& 0.383\%&5.04& 1.81 	&-0.955&-0.991&0.984&-0.757&0.884&0.809\\
& $10^5$&$10^5$&  0.295&9.54&0.0163\%& 1.33\%&6.12& 2.33 	&-0.944&-0.983&0.986&-0.749&0.891&0.823\\
& $10^5$&$10^6$&  0.351&142&0.0574\%& 2.03\%&5.93& 118 	&-0.961&-0.990&0.989&-0.938&0.942&0.891\\
& $10^6$&$10^6$&  0.415&228&0.138\%& 5.33\%&8.30& 286	&-0.956&-0.986&0.990&-0.820&0.935&0.885\\
\hline \hline
\end{tabular}
\caption{Root-mean-squared errors for source parameters, the corresponding bounds on $\lambda_g$ and $\lambda_{\mathbb{A}}$, and the correlation coefficients, for the case $\alpha=3$ and for systems with different masses in units of $M_{\odot}$. 
The top cluster uses the Ad.~LIGO $S_{n}(f)$, $\rho =10$, $\lambda_{g}$ is in units of $10^{12} \; {\rm{km}}$, $\lambda_{\mathbb{A}}$ is in units of $10^{-16}\, \rm{km}$ and $\Delta t_c$ is in msecs.
The middle cluster uses the ET $S_{n}(f)$, $\rho = 50$, $\lambda_g$ is in units of $10^{13} \; {\rm{km}}$, $\lambda_{\mathbb{A}}$ is in units of $10^{-15}\, \rm{km}$ and $\Delta t_c$ is in msecs. 
The bottom cluster uses a NGO $S_{n}(f)$, $\rho = 100$, $\lambda_g$ is in units of $10^{15} \; {\rm{km}}$, $\lambda_{\mathbb{A}}$ is in units of $10^{-10}\, \rm{km}$ and $\Delta t_c$ is in secs.
} 
\label{table-all}
\end{table}
%%%%%%%
The bounds on $\beta$ and $\zeta$ are converted into  a lower bound on $\lambda_{g}$ and and upper bound on $\lambda_{\mathbb{A}}$ in Table~\ref{table-all} for $\alpha=3$ and binary systems with different component masses. Given a gravitational-wave detection consistent with general relativity, this table says that $\lambda_{g}$ and $\lambda_{\mathbb{A}}$ would have to be larger and smaller than the numbers in the seventh and eight columns of the table respectively. In addition, this table also shows the accuracy to which standard binary parameters could be measured, such as the time of coalescence, the chirp mass and the symmetric mass ratio, as well as the correlation coefficients between parameters. Different clusters of numbers correspond to constraints with Ad.~LIGO (top), New ET (middle) and NGO (bottom -- see caption for further details). 

Although Fig.~\ref{Zeta-figs} suggests bounds on $\zeta$ of ${\cal{O}}(10^{3}-10^{5})$ for the $\alpha=3$ case, the dimensional bounds in Table~\ref{table-all} suggest a strong constraint on $\lambda_{\mathbb{A}}$. This is because in converting from $\zeta$ to $\lambda_{\mathbb{A}}$ one must divide by the $D_{3}$ distance measure. This distance is comparable to (but smaller than) the luminosity distance, and thus, the longer the graviton propagates the more sensitive the constraints are to possible Lorentz violations.  Second, notice that the accuracy to which many parameters can be determined, e.g.~$(t_c, \Delta{\mathcal M}, \Delta\eta)$, degrades with total mass because the number of observed gravitational-wave cycles decreases. Third, notice that the bound on the graviton Compton wavelength is not greatly affected by the inclusion of an additional parameter in the $\alpha=3$ case, and is comparable to the one obtained in~\cite{Will:1997bb} for LIGO. In fact, we have checked that in the absence of $\lambda_{\mathbb{A}}$ we recover Table~II in~\cite{Will:1997bb}.

\begin{figure*}[ht]
\begin{tabular}{rl}
\includegraphics[width=9cm,clip=true]{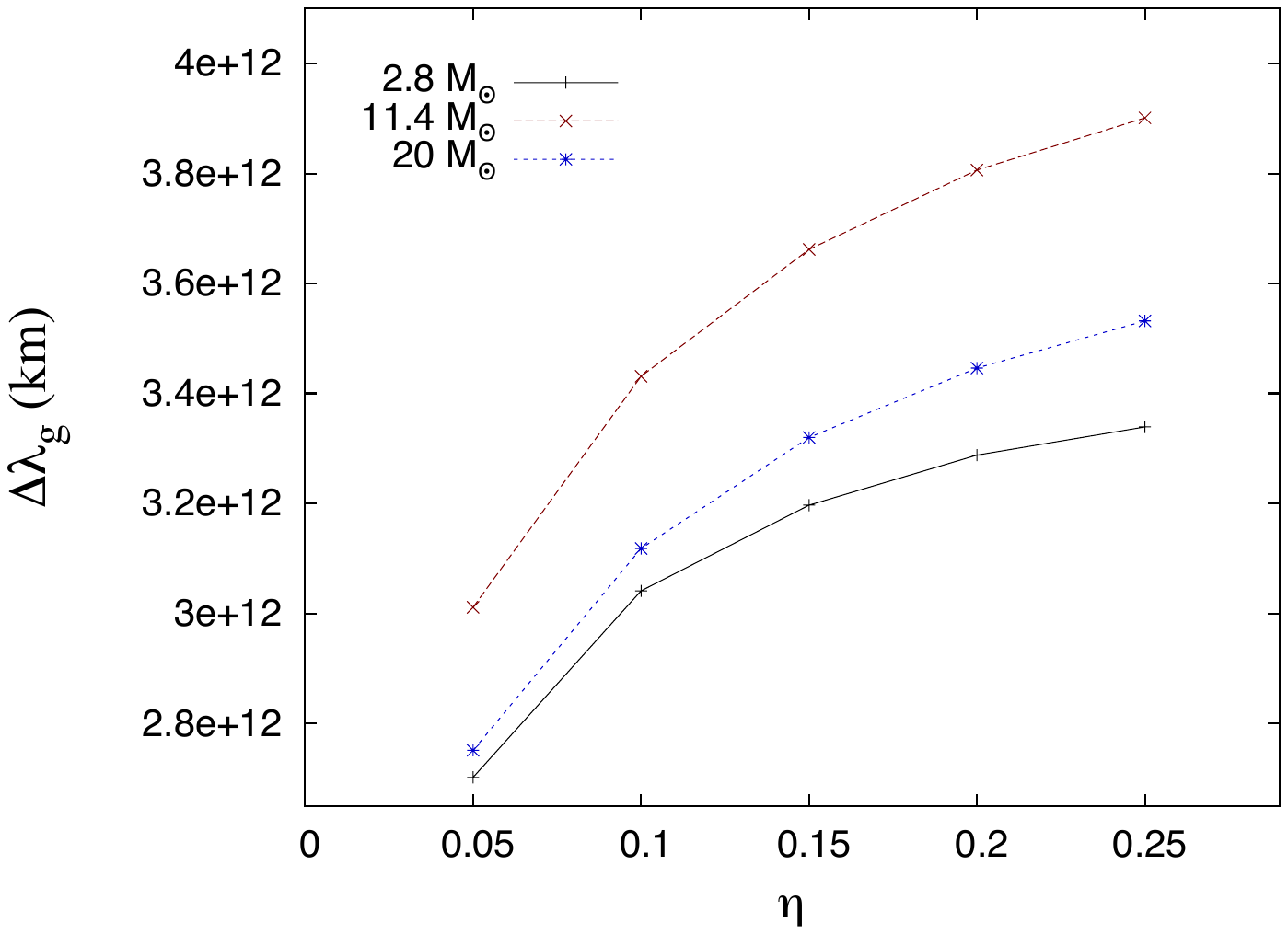} &
\includegraphics[width=9cm,clip=true]{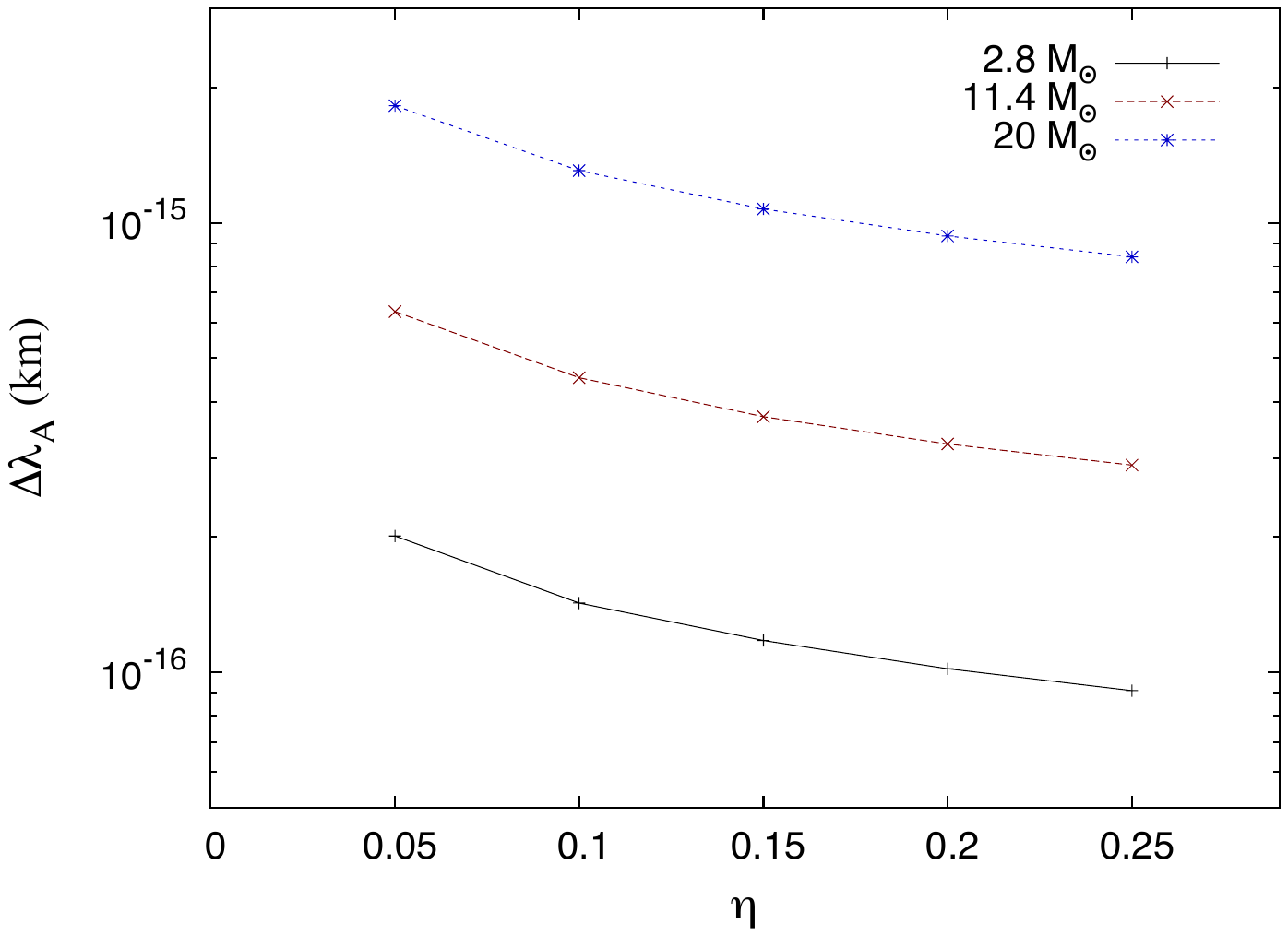} 
\end{tabular}
\caption{\label{Other-Bounds-figs} Bounds on $\lambda_g$ (left) and $\lambda_{\mathbb A}$ (right) as a function of $\eta$ for different total masses, Ad.~LIGO, $\rho = 10$ and $\alpha=3$. }
\end{figure*}
We now consider how these bounds behave as a function of the mass ratio. Figure~\ref{Other-Bounds-figs} plots the bound on the graviton Compton wavelength (left) and the Lorentz-violating Compton wavelength $\lambda_{\mathbb{A}}$ (right) as a function of $\eta$ for Ad.~LIGO and $\alpha =3$, with systems of different total mass. Notice that, in general, both bounds improve for comparable mass systems, even though the SNR is kept fixed. 

With all of this information at hand, it seems likely that gravitational-wave detection would provide useful information about Lorentz-violating graviton propagation. For example, if a Bayesian analysis were carried out, once a gravitational wave is detected, and the ppE parameters peaked around $b_{\ppE} = 2$ or $3$, this could possibly indicate the presence of some degree of Lorentz violation. Complementarily, if no deviation from general relativity is observed, then one could constrain the magnitude of $\mathbb{A}$ to interesting levels, considering that no bounds exist to date.

%%%%%%%%%%%%%%%%%%%%%%%%%%%%%
%%%%%%%%%%%%%%%%%%%%%%%%%%%%%
\section{Conclusions and Discussion}\label{conclusions}

We studied whether Lorentz symmetry-breaking in the propagation of gravitational waves could be measured with gravitational waves from non-spinning, compact binary inspirals. We considered modifications to a massive graviton dispersion relation that scale as $\mathbb{A} p^{\alpha}$, where $p$ is the graviton's momentum while $(\mathbb{A},\alpha)$ are phenomenological parameters. We found that such a modification introduces new terms in the gravitational-wave phase due to a delay in the propagation: waves emitted at low frequency, early in the inspiral, travel slightly slower than those emitted at high frequency later. This results in an offset in the relative arrival times at a detector, and thus, a frequency-dependent phase correction. We mapped these new gravitational-wave phase terms to the recently proposed ppE scheme, with ppE phase parameters $b_{\ppE} = \alpha-1$.

We then carried out a simple Fisher analysis to get a sense of the accuracy to which such dispersion relation deviations could be measured with different gravitational-wave detectors. We found that indeed, both the mass of the graviton and additional dispersion relation deviations could be constrained. For values of $\alpha>4$, there is not enough information in the waveform to produce an invertible Fisher matrix. Certain values of $\alpha$, like $\alpha = 0$ and $2$, also cannot be measured, as they become $100\%$ correlated with other system parameters. 

In deriving these bounds, we have made several approximations that force us to consider them only as rough indicators that gravitational waves can be used to constrain generic Lorentz-violation in gravitational-wave propagation. For example, we have not accounted for precession or eccentricity in the orbits, the merger phase of the inspiral, the spins of the compact objects or carried out a Bayesian analysis. We expect the inclusion of these effects to modify and possibly worsen the bounds presented above by roughly an order of magnitude, based on previous results for bounds on the mass of the graviton~\cite{Will:1997bb,Will:2004xi,Berti:2005qd,Stavridis:2009mb,Arun:2009pq,Keppel:2010qu,Yagi:2009zm}. However, the detection of $N$ gravitational waves would lead to a $\sqrt{N}$ improvement in the bounds~\cite{2011arXiv1107.3528B}, while the modeling of only the Lorentz-violating term, without including the mass of the graviton, would also increase the accuracy to which $\lambda_{\mathbb{A}}$ could me measured~\cite{Cornish:2011ys}.

Future work could concentrate on carrying out a more detailed data analysis study, using Bayesian techniques. In particular, it would be interesting to compute the evidence for a general relativity model and a modified dispersion relation model, given a signal consistent with general relativity, to see the betting-odds of the signal favoring GR over the non-GR model. A similar study was already carried out in~\cite{Cornish:2011ys}, but there a single ppE parameter was considered.  Another interesting avenue for future research would be to consider whether there are any theories (quantum-inspired or not) that predict fractional $\alpha$ powers or values of $\alpha$ different from $3$ or $4$. 

%%%%%%%%%%%%%%%%%%%%%%%%%%%%%
%%%%%%%%%%%%%%%%%%%%%%%%%%%%%
\acknowledgments
We would like to thank Leo Stein for help streamlining our Mathematica code. SM acknowledges Francesc Ferrer and K. G. Arun for helpful discussions. NY acknowledges support from NSF grant PHY-1114374, as well as support provided by the National Aeronautics and Space Administration through Einstein Postdoctoral Fellowship  Award Number PF0-110080, issued by the Chandra X-ray Observatory Center, which is operated by the Smithsonian Astrophysical Observatory for and on behalf of the National Aeronautics Space Administration under contract NAS8-03060. NY also acknowledges support from NASA grant NNX11AI49G, under sub-award 00001944. CMW and SM were supported in part by the National Science Foundation, Grant No.\ PHY 09--65133.  CMW thanks the Institut d'Astrophysique de Paris and SM thanks the MIT Kavli Institute for their hospitality during the completion of this work. 

%%%%%%%%%%%%%%%%%%%%%%%%%%%%%
%%%%%%%%%%%%%%%%%%%%%%%%%%%%%
\bibliography{master}

\begin{thebibliography}{74}
\expandafter\ifx\csname natexlab\endcsname\relax\def\natexlab#1{#1}\fi
\expandafter\ifx\csname bibnamefont\endcsname\relax
  \def\bibnamefont#1{#1}\fi
\expandafter\ifx\csname bibfnamefont\endcsname\relax
  \def\bibfnamefont#1{#1}\fi
\expandafter\ifx\csname citenamefont\endcsname\relax
  \def\citenamefont#1{#1}\fi
\expandafter\ifx\csname url\endcsname\relax
  \def\url#1{\texttt{#1}}\fi
\expandafter\ifx\csname urlprefix\endcsname\relax\def\urlprefix{URL }\fi
\providecommand{\bibinfo}[2]{#2}
\providecommand{\eprint}[2][]{\url{#2}}

\bibitem[{\citenamefont{Bird et~al.}(1995)}]{Bird:1994uy}
\bibinfo{author}{\bibfnamefont{D.~J.} \bibnamefont{Bird}} \bibnamefont{et~al.},
  \bibinfo{journal}{Astrophys. J.} \textbf{\bibinfo{volume}{441}},
  \bibinfo{pages}{144} (\bibinfo{year}{1995}).

\bibitem[{lig()}]{ligo}
\emph{\bibinfo{title}{{LIGO}}}, \bibinfo{note}{{\tt www.ligo.caltech.edu}}.

\bibitem[{\citenamefont{Abbott et~al.}(2009)}]{Abbott:2007kv}
\bibinfo{author}{\bibfnamefont{B.}~\bibnamefont{Abbott}} \bibnamefont{et~al.},
  \bibinfo{journal}{Rept. Prog. Phys.} \textbf{\bibinfo{volume}{72}},
  \bibinfo{pages}{076901} (\bibinfo{year}{2009}), \eprint{0711.3041}.

\bibitem[{\citenamefont{{Harry} and {the LIGO Scientific
  Collaboration}}(2010)}]{2010CQGra..27h4006H}
\bibinfo{author}{\bibfnamefont{G.~M.} \bibnamefont{{Harry}}} \bibnamefont{and}
  \bibinfo{author}{\bibnamefont{{the LIGO Scientific Collaboration}}},
  \bibinfo{journal}{Classical and Quantum Gravity}
  \textbf{\bibinfo{volume}{27}}, \bibinfo{pages}{084006}
  (\bibinfo{year}{2010}).

\bibitem[{vir()}]{virgo}
\emph{\bibinfo{title}{{VIRGO}}}, \bibinfo{note}{{\tt www.virgo.infn.it}}.

\bibitem[{et()}]{et}
\emph{\bibinfo{title}{{Einstein Telescope}}}, \bibinfo{note}{{\tt
  www.et-gw.eu}}.

\bibitem[{\citenamefont{Punturo et~al.}(2010)\citenamefont{Punturo, Abernathy,
  Acernese, Allen, Andersson et~al.}}]{Punturo:2010zz}
\bibinfo{author}{\bibfnamefont{M.}~\bibnamefont{Punturo}},
  \bibinfo{author}{\bibfnamefont{M.}~\bibnamefont{Abernathy}},
  \bibinfo{author}{\bibfnamefont{F.}~\bibnamefont{Acernese}},
  \bibinfo{author}{\bibfnamefont{B.}~\bibnamefont{Allen}},
  \bibinfo{author}{\bibfnamefont{N.}~\bibnamefont{Andersson}},
  \bibnamefont{et~al.}, \bibinfo{journal}{Class.Quant.Grav.}
  \textbf{\bibinfo{volume}{27}}, \bibinfo{pages}{194002}
  (\bibinfo{year}{2010}).

\bibitem[{lis()}]{lisa}
\emph{\bibinfo{title}{{LISA}}}, \bibinfo{note}{{\tt www.esa.int/science/lisa},
  {\tt lisa.jpl.nasa.gov}}.

\bibitem[{\citenamefont{{Prince}}(2003)}]{Prince:2003aa}
\bibinfo{author}{\bibfnamefont{T.}~\bibnamefont{{Prince}}},
  \bibinfo{journal}{American Astronomical Society Meeting}
  \textbf{\bibinfo{volume}{202}}, \bibinfo{pages}{3701} (\bibinfo{year}{2003}).

\bibitem[{\citenamefont{Will}(1993)}]{Will:1993ns}
\bibinfo{author}{\bibfnamefont{C.~M.} \bibnamefont{Will}},
  \emph{\bibinfo{title}{Theory and experiment in gravitational physics}}
  (\bibinfo{publisher}{Cambridge University Press},
  \bibinfo{address}{Cambridge, UK}, \bibinfo{year}{1993}).

\bibitem[{\citenamefont{Will}(1998)}]{Will:1997bb}
\bibinfo{author}{\bibfnamefont{C.~M.} \bibnamefont{Will}},
  \bibinfo{journal}{Phys. Rev.} \textbf{\bibinfo{volume}{D57}},
  \bibinfo{pages}{2061} (\bibinfo{year}{1998}), \eprint{gr-qc/9709011}.

\bibitem[{\citenamefont{Yunes and Pretorius}(2009)}]{Yunes:2009ke}
\bibinfo{author}{\bibfnamefont{N.}~\bibnamefont{Yunes}} \bibnamefont{and}
  \bibinfo{author}{\bibfnamefont{F.}~\bibnamefont{Pretorius}},
  \bibinfo{journal}{Phys.Rev.} \textbf{\bibinfo{volume}{D80}},
  \bibinfo{pages}{122003} (\bibinfo{year}{2009}), \eprint{0909.3328}.

\bibitem[{\citenamefont{Cornish et~al.}(2011)\citenamefont{Cornish, Sampson,
  Yunes, and Pretorius}}]{Cornish:2011ys}
\bibinfo{author}{\bibfnamefont{N.}~\bibnamefont{Cornish}},
  \bibinfo{author}{\bibfnamefont{L.}~\bibnamefont{Sampson}},
  \bibinfo{author}{\bibfnamefont{N.}~\bibnamefont{Yunes}}, \bibnamefont{and}
  \bibinfo{author}{\bibfnamefont{F.}~\bibnamefont{Pretorius}}
  (\bibinfo{year}{2011}), \bibinfo{note}{* Temporary entry *},
  \eprint{1105.2088}.

\bibitem[{\citenamefont{Will and Yunes}(2004)}]{Will:2004xi}
\bibinfo{author}{\bibfnamefont{C.~M.} \bibnamefont{Will}} \bibnamefont{and}
  \bibinfo{author}{\bibfnamefont{N.}~\bibnamefont{Yunes}},
  \bibinfo{journal}{Class. Quant. Grav.} \textbf{\bibinfo{volume}{21}},
  \bibinfo{pages}{4367} (\bibinfo{year}{2004}), \eprint{gr-qc/0403100}.

\bibitem[{\citenamefont{Berti et~al.}(2005{\natexlab{a}})\citenamefont{Berti,
  Buonanno, and Will}}]{Berti:2004bd}
\bibinfo{author}{\bibfnamefont{E.}~\bibnamefont{Berti}},
  \bibinfo{author}{\bibfnamefont{A.}~\bibnamefont{Buonanno}}, \bibnamefont{and}
  \bibinfo{author}{\bibfnamefont{C.~M.} \bibnamefont{Will}},
  \bibinfo{journal}{Phys. Rev.} \textbf{\bibinfo{volume}{D71}},
  \bibinfo{pages}{084025} (\bibinfo{year}{2005}{\natexlab{a}}),
  \eprint{gr-qc/0411129}.

\bibitem[{\citenamefont{Arun and Will}(2009)}]{Arun:2009pq}
\bibinfo{author}{\bibfnamefont{K.~G.} \bibnamefont{Arun}} \bibnamefont{and}
  \bibinfo{author}{\bibfnamefont{C.~M.} \bibnamefont{Will}}
  (\bibinfo{year}{2009}), \eprint{0904.1190}.

\bibitem[{\citenamefont{{Keppel} and {Ajith}}(2010)}]{2010PhRvD..82l2001K}
\bibinfo{author}{\bibfnamefont{D.}~\bibnamefont{{Keppel}}} \bibnamefont{and}
  \bibinfo{author}{\bibfnamefont{P.}~\bibnamefont{{Ajith}}},
  \bibinfo{journal}{\prd} \textbf{\bibinfo{volume}{82}},
  \bibinfo{pages}{122001} (\bibinfo{year}{2010}), \eprint{1004.0284}.

\bibitem[{\citenamefont{{Yagi} and {Tanaka}}(2010)}]{2010PhRvD..81f4008Y}
\bibinfo{author}{\bibfnamefont{K.}~\bibnamefont{{Yagi}}} \bibnamefont{and}
  \bibinfo{author}{\bibfnamefont{T.}~\bibnamefont{{Tanaka}}},
  \bibinfo{journal}{\prd} \textbf{\bibinfo{volume}{81}},
  \bibinfo{pages}{064008} (\bibinfo{year}{2010}), \eprint{0906.4269}.

\bibitem[{\citenamefont{{Berti} et~al.}(2011)\citenamefont{{Berti}, {Gair}, and
  {Sesana}}}]{2011arXiv1107.3528B}
\bibinfo{author}{\bibfnamefont{E.}~\bibnamefont{{Berti}}},
  \bibinfo{author}{\bibfnamefont{J.}~\bibnamefont{{Gair}}}, \bibnamefont{and}
  \bibinfo{author}{\bibfnamefont{A.}~\bibnamefont{{Sesana}}},
  \bibinfo{journal}{ArXiv e-prints}  (\bibinfo{year}{2011}),
  \eprint{1107.3528}.

\bibitem[{\citenamefont{Biller et~al.}(1999)\citenamefont{Biller, Breslin,
  Buckley, Catanese, Carson et~al.}}]{Biller:1998hg}
\bibinfo{author}{\bibfnamefont{S.}~\bibnamefont{Biller}},
  \bibinfo{author}{\bibfnamefont{A.}~\bibnamefont{Breslin}},
  \bibinfo{author}{\bibfnamefont{J.}~\bibnamefont{Buckley}},
  \bibinfo{author}{\bibfnamefont{M.}~\bibnamefont{Catanese}},
  \bibinfo{author}{\bibfnamefont{M.}~\bibnamefont{Carson}},
  \bibnamefont{et~al.}, \bibinfo{journal}{Phys.Rev.Lett.}
  \textbf{\bibinfo{volume}{83}}, \bibinfo{pages}{2108} (\bibinfo{year}{1999}),
  \eprint{gr-qc/9810044}.

\bibitem[{\citenamefont{{Blas} and {Sanctuary}}(2011)}]{2011PhRvD..84f4004B}
\bibinfo{author}{\bibfnamefont{D.}~\bibnamefont{{Blas}}} \bibnamefont{and}
  \bibinfo{author}{\bibfnamefont{H.}~\bibnamefont{{Sanctuary}}},
  \bibinfo{journal}{\prd} \textbf{\bibinfo{volume}{84}}, \bibinfo{eid}{064004}
  (\bibinfo{year}{2011}), \eprint{1105.5149}.

\bibitem[{\citenamefont{{Jacobson} and
  {Mattingly}}(2004)}]{2004PhRvD..70b4003J}
\bibinfo{author}{\bibfnamefont{T.}~\bibnamefont{{Jacobson}}} \bibnamefont{and}
  \bibinfo{author}{\bibfnamefont{D.}~\bibnamefont{{Mattingly}}},
  \bibinfo{journal}{\prd} \textbf{\bibinfo{volume}{70}}, \bibinfo{eid}{024003}
  (\bibinfo{year}{2004}), \eprint{arXiv:gr-qc/0402005}.

\bibitem[{\citenamefont{{Foster}}(2007)}]{2007PhRvD..76h4033F}
\bibinfo{author}{\bibfnamefont{B.~Z.} \bibnamefont{{Foster}}},
  \bibinfo{journal}{\prd} \textbf{\bibinfo{volume}{76}}, \bibinfo{eid}{084033}
  (\bibinfo{year}{2007}), \eprint{0706.0704}.

\bibitem[{\citenamefont{{Sagi}}(2010)}]{2010PhRvD..81f4031S}
\bibinfo{author}{\bibfnamefont{E.}~\bibnamefont{{Sagi}}},
  \bibinfo{journal}{\prd} \textbf{\bibinfo{volume}{81}}, \bibinfo{eid}{064031}
  (\bibinfo{year}{2010}), \eprint{1001.1555}.

\bibitem[{\citenamefont{{Sagi}}(2009)}]{2009PhRvD..80d4032S}
\bibinfo{author}{\bibfnamefont{E.}~\bibnamefont{{Sagi}}},
  \bibinfo{journal}{\prd} \textbf{\bibinfo{volume}{80}}, \bibinfo{eid}{044032}
  (\bibinfo{year}{2009}), \eprint{0905.4001}.

\bibitem[{\citenamefont{{Hohmann}}(2011)}]{2011arXiv1105.2555H}
\bibinfo{author}{\bibfnamefont{M.}~\bibnamefont{{Hohmann}}},
  \bibinfo{journal}{ArXiv e-prints}  (\bibinfo{year}{2011}),
  \eprint{1105.2555}.

\bibitem[{\citenamefont{{Radicella} and
  {Tartaglia}}(2010)}]{2010AIPC.1241.1128R}
\bibinfo{author}{\bibfnamefont{N.}~\bibnamefont{{Radicella}}} \bibnamefont{and}
  \bibinfo{author}{\bibfnamefont{A.}~\bibnamefont{{Tartaglia}}}, in
  \emph{\bibinfo{booktitle}{American Institute of Physics Conference Series}},
  edited by \bibinfo{editor}{\bibnamefont{{J.-M.~Alimi \& A.~Fu{\"o}zfa}}}
  (\bibinfo{year}{2010}), vol. \bibinfo{volume}{1241} of
  \emph{\bibinfo{series}{American Institute of Physics Conference Series}}, pp.
  \bibinfo{pages}{1128--1133}, \eprint{0911.3365}.

\bibitem[{\citenamefont{{Bellucci} et~al.}(2009)\citenamefont{{Bellucci},
  {Capozziello}, {de Laurentis}, and {Faraoni}}}]{2009PhRvD..79j4004B}
\bibinfo{author}{\bibfnamefont{S.}~\bibnamefont{{Bellucci}}},
  \bibinfo{author}{\bibfnamefont{S.}~\bibnamefont{{Capozziello}}},
  \bibinfo{author}{\bibfnamefont{M.}~\bibnamefont{{de Laurentis}}},
  \bibnamefont{and}
  \bibinfo{author}{\bibfnamefont{V.}~\bibnamefont{{Faraoni}}},
  \bibinfo{journal}{\prd} \textbf{\bibinfo{volume}{79}}, \bibinfo{eid}{104004}
  (\bibinfo{year}{2009}), \eprint{0812.1348}.

\bibitem[{\citenamefont{{M{\"u}ller} et~al.}(2011)\citenamefont{{M{\"u}ller},
  {Alves}, and {de Araujo}}}]{2011arXiv1103.3439M}
\bibinfo{author}{\bibfnamefont{D.}~\bibnamefont{{M{\"u}ller}}},
  \bibinfo{author}{\bibfnamefont{M.~E.~S.} \bibnamefont{{Alves}}},
  \bibnamefont{and} \bibinfo{author}{\bibfnamefont{J.~C.~N.} \bibnamefont{{de
  Araujo}}}, \bibinfo{journal}{ArXiv e-prints}  (\bibinfo{year}{2011}),
  \eprint{1103.3439}.

\bibitem[{\citenamefont{{Pani} et~al.}(2011)\citenamefont{{Pani}, {Berti},
  {Cardoso}, and {Read}}}]{2011PhRvD..84j4035P}
\bibinfo{author}{\bibfnamefont{P.}~\bibnamefont{{Pani}}},
  \bibinfo{author}{\bibfnamefont{E.}~\bibnamefont{{Berti}}},
  \bibinfo{author}{\bibfnamefont{V.}~\bibnamefont{{Cardoso}}},
  \bibnamefont{and} \bibinfo{author}{\bibfnamefont{J.}~\bibnamefont{{Read}}},
  \bibinfo{journal}{\prd} \textbf{\bibinfo{volume}{84}}, \bibinfo{eid}{104035}
  (\bibinfo{year}{2011}), \eprint{1109.0928}.

\bibitem[{\citenamefont{{de Rham} and {Tolley}}(2006)}]{2006JCAP...02..003D}
\bibinfo{author}{\bibfnamefont{C.}~\bibnamefont{{de Rham}}} \bibnamefont{and}
  \bibinfo{author}{\bibfnamefont{A.~J.} \bibnamefont{{Tolley}}},
  \bibinfo{journal}{J. Cosmol. Astropart. Phys.} \textbf{\bibinfo{volume}{2}},
  \bibinfo{pages}{3} (\bibinfo{year}{2006}), \eprint{arXiv:hep-th/0511138}.

\bibitem[{\citenamefont{{Yagi} et~al.}(2011)\citenamefont{{Yagi}, {Stein},
  {Yunes}, and {Tanaka}}}]{2011arXiv1110.5950Y}
\bibinfo{author}{\bibfnamefont{K.}~\bibnamefont{{Yagi}}},
  \bibinfo{author}{\bibfnamefont{L.~C.} \bibnamefont{{Stein}}},
  \bibinfo{author}{\bibfnamefont{N.}~\bibnamefont{{Yunes}}}, \bibnamefont{and}
  \bibinfo{author}{\bibfnamefont{T.}~\bibnamefont{{Tanaka}}},
  \bibinfo{journal}{ArXiv e-prints}  (\bibinfo{year}{2011}),
  \eprint{1110.5950}.

\bibitem[{\citenamefont{{Bekenstein}}(2004)}]{2004PhRvD..70h3509B}
\bibinfo{author}{\bibfnamefont{J.~D.} \bibnamefont{{Bekenstein}}},
  \bibinfo{journal}{\prd} \textbf{\bibinfo{volume}{70}}, \bibinfo{eid}{083509}
  (\bibinfo{year}{2004}), \eprint{arXiv:astro-ph/0403694}.

\bibitem[{\citenamefont{{Amelino-Camelia}}(2001)}]{2001PhLB..510..255A}
\bibinfo{author}{\bibfnamefont{G.}~\bibnamefont{{Amelino-Camelia}}},
  \bibinfo{journal}{Physics Letters B} \textbf{\bibinfo{volume}{510}},
  \bibinfo{pages}{255} (\bibinfo{year}{2001}), \eprint{arXiv:hep-th/0012238}.

\bibitem[{\citenamefont{{Magueijo} and {Smolin}}(2002)}]{2002PhRvL..88s0403M}
\bibinfo{author}{\bibfnamefont{J.}~\bibnamefont{{Magueijo}}} \bibnamefont{and}
  \bibinfo{author}{\bibfnamefont{L.}~\bibnamefont{{Smolin}}},
  \bibinfo{journal}{Physical Review Letters} \textbf{\bibinfo{volume}{88}},
  \bibinfo{pages}{190403} (\bibinfo{year}{2002}),
  \eprint{arXiv:hep-th/0112090}.

\bibitem[{\citenamefont{Amelino-Camelia}(2002)}]{AmelinoCamelia:2002wr}
\bibinfo{author}{\bibfnamefont{G.}~\bibnamefont{Amelino-Camelia}},
  \bibinfo{journal}{Nature} \textbf{\bibinfo{volume}{418}}, \bibinfo{pages}{34}
  (\bibinfo{year}{2002}), \eprint{gr-qc/0207049}.

\bibitem[{\citenamefont{{Amelino-Camelia}}(2010)}]{2010arXiv1003.3942A}
\bibinfo{author}{\bibfnamefont{G.}~\bibnamefont{{Amelino-Camelia}}},
  \bibinfo{journal}{ArXiv e-prints}  (\bibinfo{year}{2010}),
  \eprint{1003.3942}.

\bibitem[{\citenamefont{{Sefiedgar} et~al.}(2011)\citenamefont{{Sefiedgar},
  {Nozari}, and {Sepangi}}}]{2011PhLB..696..119S}
\bibinfo{author}{\bibfnamefont{A.~S.} \bibnamefont{{Sefiedgar}}},
  \bibinfo{author}{\bibfnamefont{K.}~\bibnamefont{{Nozari}}}, \bibnamefont{and}
  \bibinfo{author}{\bibfnamefont{H.~R.} \bibnamefont{{Sepangi}}},
  \bibinfo{journal}{Physics Letters B} \textbf{\bibinfo{volume}{696}},
  \bibinfo{pages}{119} (\bibinfo{year}{2011}), \eprint{1012.1406}.

\bibitem[{\citenamefont{Horava}(2009{\natexlab{a}})}]{Horava:2008ih}
\bibinfo{author}{\bibfnamefont{P.}~\bibnamefont{Horava}},
  \bibinfo{journal}{JHEP} \textbf{\bibinfo{volume}{0903}}, \bibinfo{pages}{020}
  (\bibinfo{year}{2009}{\natexlab{a}}), \eprint{0812.4287}.

\bibitem[{\citenamefont{Horava}(2009{\natexlab{b}})}]{Horava:2009uw}
\bibinfo{author}{\bibfnamefont{P.}~\bibnamefont{Horava}},
  \bibinfo{journal}{Phys.Rev.} \textbf{\bibinfo{volume}{D79}},
  \bibinfo{pages}{084008} (\bibinfo{year}{2009}{\natexlab{b}}),
  \eprint{0901.3775}.

\bibitem[{\citenamefont{{Vacaru}}(2010)}]{2010arXiv1010.5457V}
\bibinfo{author}{\bibfnamefont{S.~I.} \bibnamefont{{Vacaru}}},
  \bibinfo{journal}{ArXiv e-prints}  (\bibinfo{year}{2010}),
  \eprint{1010.5457}.

\bibitem[{\citenamefont{Blas and Sanctuary}(2011)}]{Blas:2011zd}
\bibinfo{author}{\bibfnamefont{D.}~\bibnamefont{Blas}} \bibnamefont{and}
  \bibinfo{author}{\bibfnamefont{H.}~\bibnamefont{Sanctuary}},
  \bibinfo{journal}{Phys.Rev.} \textbf{\bibinfo{volume}{D84}},
  \bibinfo{pages}{064004} (\bibinfo{year}{2011}), \eprint{1105.5149}.

\bibitem[{\citenamefont{{Garattini}}(2011)}]{2011arXiv1102.0117G}
\bibinfo{author}{\bibfnamefont{R.}~\bibnamefont{{Garattini}}},
  \bibinfo{journal}{ArXiv e-prints}  (\bibinfo{year}{2011}),
  \eprint{1102.0117}.

\bibitem[{\citenamefont{Garattini and
  Mandanici}(2011{\natexlab{a}})}]{Garattini:2011kp}
\bibinfo{author}{\bibfnamefont{R.}~\bibnamefont{Garattini}} \bibnamefont{and}
  \bibinfo{author}{\bibfnamefont{G.}~\bibnamefont{Mandanici}},
  \bibinfo{journal}{Phys.Rev.} \textbf{\bibinfo{volume}{D83}},
  \bibinfo{pages}{084021} (\bibinfo{year}{2011}{\natexlab{a}}),
  \eprint{1102.3803}.

\bibitem[{\citenamefont{Garattini and
  Mandanici}(2011{\natexlab{b}})}]{Garattini:2011hy}
\bibinfo{author}{\bibfnamefont{R.}~\bibnamefont{Garattini}} \bibnamefont{and}
  \bibinfo{author}{\bibfnamefont{G.}~\bibnamefont{Mandanici}}
  (\bibinfo{year}{2011}{\natexlab{b}}), \eprint{1109.6563}.

\bibitem[{\citenamefont{Berezhiani et~al.}(2007)\citenamefont{Berezhiani,
  Comelli, Nesti, and Pilo}}]{Berezhiani:2007zf}
\bibinfo{author}{\bibfnamefont{Z.}~\bibnamefont{Berezhiani}},
  \bibinfo{author}{\bibfnamefont{D.}~\bibnamefont{Comelli}},
  \bibinfo{author}{\bibfnamefont{F.}~\bibnamefont{Nesti}}, \bibnamefont{and}
  \bibinfo{author}{\bibfnamefont{L.}~\bibnamefont{Pilo}},
  \bibinfo{journal}{Phys.Rev.Lett.} \textbf{\bibinfo{volume}{99}},
  \bibinfo{pages}{131101} (\bibinfo{year}{2007}), \eprint{hep-th/0703264}.

\bibitem[{\citenamefont{Berezhiani et~al.}(2008)\citenamefont{Berezhiani,
  Comelli, Nesti, and Pilo}}]{Berezhiani:2008nr}
\bibinfo{author}{\bibfnamefont{Z.}~\bibnamefont{Berezhiani}},
  \bibinfo{author}{\bibfnamefont{D.}~\bibnamefont{Comelli}},
  \bibinfo{author}{\bibfnamefont{F.}~\bibnamefont{Nesti}}, \bibnamefont{and}
  \bibinfo{author}{\bibfnamefont{L.}~\bibnamefont{Pilo}},
  \bibinfo{journal}{JHEP} \textbf{\bibinfo{volume}{0807}}, \bibinfo{pages}{130}
  (\bibinfo{year}{2008}), \eprint{0803.1687}.

\bibitem[{\citenamefont{{Bojowald} and {Hossain}}(2008)}]{2008PhRvD..77b3508B}
\bibinfo{author}{\bibfnamefont{M.}~\bibnamefont{{Bojowald}}} \bibnamefont{and}
  \bibinfo{author}{\bibfnamefont{G.~M.} \bibnamefont{{Hossain}}},
  \bibinfo{journal}{\prd} \textbf{\bibinfo{volume}{77}},
  \bibinfo{pages}{023508} (\bibinfo{year}{2008}), \eprint{0709.2365}.

\bibitem[{\citenamefont{{Chouha} and
  {Brandenberger}}(2005)}]{2005hep.th....8119C}
\bibinfo{author}{\bibfnamefont{P.~R.} \bibnamefont{{Chouha}}} \bibnamefont{and}
  \bibinfo{author}{\bibfnamefont{R.~H.} \bibnamefont{{Brandenberger}}},
  \bibinfo{journal}{ArXiv High Energy Physics - Theory e-prints}
  (\bibinfo{year}{2005}), \eprint{arXiv:hep-th/0508119}.

\bibitem[{\citenamefont{{Szabo}}(2010)}]{2010GReGr..42....1S}
\bibinfo{author}{\bibfnamefont{R.~J.} \bibnamefont{{Szabo}}},
  \bibinfo{journal}{General Relativity and Gravitation}
  \textbf{\bibinfo{volume}{42}}, \bibinfo{pages}{1} (\bibinfo{year}{2010}),
  \eprint{0906.2913}.

\bibitem[{\citenamefont{{Collins} et~al.}(2004)\citenamefont{{Collins},
  {Perez}, {Sudarsky}, {Urrutia}, and {Vucetich}}}]{2004PhRvL..93s1301C}
\bibinfo{author}{\bibfnamefont{J.}~\bibnamefont{{Collins}}},
  \bibinfo{author}{\bibfnamefont{A.}~\bibnamefont{{Perez}}},
  \bibinfo{author}{\bibfnamefont{D.}~\bibnamefont{{Sudarsky}}},
  \bibinfo{author}{\bibfnamefont{L.}~\bibnamefont{{Urrutia}}},
  \bibnamefont{and}
  \bibinfo{author}{\bibfnamefont{H.}~\bibnamefont{{Vucetich}}},
  \bibinfo{journal}{Physical Review Letters} \textbf{\bibinfo{volume}{93}},
  \bibinfo{pages}{191301} (\bibinfo{year}{2004}), \eprint{arXiv:gr-qc/0403053}.

\bibitem[{\citenamefont{{Collins} et~al.}(2006)\citenamefont{{Collins},
  {Perez}, and {Sudarsky}}}]{2006hep.th....3002C}
\bibinfo{author}{\bibfnamefont{J.}~\bibnamefont{{Collins}}},
  \bibinfo{author}{\bibfnamefont{A.}~\bibnamefont{{Perez}}}, \bibnamefont{and}
  \bibinfo{author}{\bibfnamefont{D.}~\bibnamefont{{Sudarsky}}},
  \bibinfo{journal}{ArXiv High Energy Physics - Theory e-prints}
  (\bibinfo{year}{2006}), \eprint{arXiv:hep-th/0603002}.

\bibitem[{\citenamefont{{Gambini} et~al.}(2011)\citenamefont{{Gambini},
  {Rastgoo}, and {Pullin}}}]{2011arXiv1106.1417G}
\bibinfo{author}{\bibfnamefont{R.}~\bibnamefont{{Gambini}}},
  \bibinfo{author}{\bibfnamefont{S.}~\bibnamefont{{Rastgoo}}},
  \bibnamefont{and} \bibinfo{author}{\bibfnamefont{J.}~\bibnamefont{{Pullin}}},
  \bibinfo{journal}{ArXiv e-prints}  (\bibinfo{year}{2011}),
  \eprint{1106.1417}.

\bibitem[{\citenamefont{Droz et~al.}(1999)\citenamefont{Droz, Knapp, Poisson,
  and Owen}}]{Droz:1999qx}
\bibinfo{author}{\bibfnamefont{S.}~\bibnamefont{Droz}},
  \bibinfo{author}{\bibfnamefont{D.~J.} \bibnamefont{Knapp}},
  \bibinfo{author}{\bibfnamefont{E.}~\bibnamefont{Poisson}}, \bibnamefont{and}
  \bibinfo{author}{\bibfnamefont{B.~J.} \bibnamefont{Owen}},
  \bibinfo{journal}{Phys.Rev.} \textbf{\bibinfo{volume}{D59}},
  \bibinfo{pages}{124016} (\bibinfo{year}{1999}), \eprint{gr-qc/9901076}.

\bibitem[{\citenamefont{Yunes et~al.}(2009)\citenamefont{Yunes, Arun, Berti,
  and Will}}]{Yunes:2009yz}
\bibinfo{author}{\bibfnamefont{N.}~\bibnamefont{Yunes}},
  \bibinfo{author}{\bibfnamefont{K.}~\bibnamefont{Arun}},
  \bibinfo{author}{\bibfnamefont{E.}~\bibnamefont{Berti}}, \bibnamefont{and}
  \bibinfo{author}{\bibfnamefont{C.~M.} \bibnamefont{Will}},
  \bibinfo{journal}{Phys.Rev.} \textbf{\bibinfo{volume}{D80}},
  \bibinfo{pages}{084001} (\bibinfo{year}{2009}), \eprint{0906.0313}.

\bibitem[{\citenamefont{{Buonanno} et~al.}(2009)\citenamefont{{Buonanno},
  {Iyer}, {Ochsner}, {Pan}, and {Sathyaprakash}}}]{2009PhRvD..80h4043B}
\bibinfo{author}{\bibfnamefont{A.}~\bibnamefont{{Buonanno}}},
  \bibinfo{author}{\bibfnamefont{B.~R.} \bibnamefont{{Iyer}}},
  \bibinfo{author}{\bibfnamefont{E.}~\bibnamefont{{Ochsner}}},
  \bibinfo{author}{\bibfnamefont{Y.}~\bibnamefont{{Pan}}}, \bibnamefont{and}
  \bibinfo{author}{\bibfnamefont{B.~S.} \bibnamefont{{Sathyaprakash}}},
  \bibinfo{journal}{\prd} \textbf{\bibinfo{volume}{80}},
  \bibinfo{pages}{084043} (\bibinfo{year}{2009}), \eprint{0907.0700}.

\bibitem[{\citenamefont{{Will}}(1971)}]{1971ApJ...163..611W}
\bibinfo{author}{\bibfnamefont{C.~M.} \bibnamefont{{Will}}},
  \bibinfo{journal}{\apj} \textbf{\bibinfo{volume}{163}}, \bibinfo{pages}{611}
  (\bibinfo{year}{1971}).

\bibitem[{\citenamefont{{Nordtvedt} and {Will}}(1972)}]{1972ApJ...177..775N}
\bibinfo{author}{\bibfnamefont{K.~J.} \bibnamefont{{Nordtvedt}}}
  \bibnamefont{and} \bibinfo{author}{\bibfnamefont{C.~M.}
  \bibnamefont{{Will}}}, \bibinfo{journal}{\apj}
  \textbf{\bibinfo{volume}{177}}, \bibinfo{pages}{775} (\bibinfo{year}{1972}).

\bibitem[{\citenamefont{{Will} and {Nordtvedt}}(1972)}]{1972ApJ...177..757W}
\bibinfo{author}{\bibfnamefont{C.~M.} \bibnamefont{{Will}}} \bibnamefont{and}
  \bibinfo{author}{\bibfnamefont{K.~J.} \bibnamefont{{Nordtvedt}}},
  \bibinfo{journal}{\apj} \textbf{\bibinfo{volume}{177}}, \bibinfo{pages}{757}
  (\bibinfo{year}{1972}).

\bibitem[{\citenamefont{{Will}}(1973)}]{1973ApJ...185...31W}
\bibinfo{author}{\bibfnamefont{C.~M.} \bibnamefont{{Will}}},
  \bibinfo{journal}{\apj} \textbf{\bibinfo{volume}{185}}, \bibinfo{pages}{31}
  (\bibinfo{year}{1973}).

\bibitem[{\citenamefont{Will}(2006)}]{lrr-2006-3}
\bibinfo{author}{\bibfnamefont{C.~M.} \bibnamefont{Will}},
  \bibinfo{journal}{Living Reviews in Relativity} \textbf{\bibinfo{volume}{9}}
  (\bibinfo{year}{2006}), \eprint{gr-qc/0510072},
  \urlprefix\url{http://www.livingreviews.org/lrr-2006-3}.

\bibitem[{\citenamefont{Vigeland et~al.}(2011)\citenamefont{Vigeland, Yunes,
  and Stein}}]{Vigeland:2011ji}
\bibinfo{author}{\bibfnamefont{S.}~\bibnamefont{Vigeland}},
  \bibinfo{author}{\bibfnamefont{N.}~\bibnamefont{Yunes}}, \bibnamefont{and}
  \bibinfo{author}{\bibfnamefont{L.}~\bibnamefont{Stein}},
  \bibinfo{journal}{Phys.Rev.} \textbf{\bibinfo{volume}{D83}},
  \bibinfo{pages}{104027} (\bibinfo{year}{2011}), \eprint{1102.3706}.

\bibitem[{\citenamefont{{Cutler} and {Flanagan}}(1994)}]{cutflan94}
\bibinfo{author}{\bibfnamefont{C.}~\bibnamefont{{Cutler}}} \bibnamefont{and}
  \bibinfo{author}{\bibfnamefont{{\'E}.~E.} \bibnamefont{{Flanagan}}},
  \bibinfo{journal}{\prd} \textbf{\bibinfo{volume}{49}}, \bibinfo{pages}{2658}
  (\bibinfo{year}{1994}), \eprint{arXiv:gr-qc/9402014}.

\bibitem[{\citenamefont{Finn and Chernoff}(1993)}]{finn}
\bibinfo{author}{\bibfnamefont{L.~S.} \bibnamefont{Finn}} \bibnamefont{and}
  \bibinfo{author}{\bibfnamefont{D.~F.} \bibnamefont{Chernoff}},
  \bibinfo{journal}{Phys. Rev.} \textbf{\bibinfo{volume}{D47}},
  \bibinfo{pages}{2198} (\bibinfo{year}{1993}), \eprint{gr-qc/9301003}.

\bibitem[{\citenamefont{Poisson and Will}(1995)}]{Poisson:1995ef}
\bibinfo{author}{\bibfnamefont{E.}~\bibnamefont{Poisson}} \bibnamefont{and}
  \bibinfo{author}{\bibfnamefont{C.~M.} \bibnamefont{Will}},
  \bibinfo{journal}{Phys. Rev.} \textbf{\bibinfo{volume}{D52}},
  \bibinfo{pages}{848} (\bibinfo{year}{1995}), \eprint{gr-qc/9502040}.

\bibitem[{\citenamefont{Mishra et~al.}(2010)\citenamefont{Mishra, Arun, Iyer,
  and Sathyaprakash}}]{Mishra:2010tp}
\bibinfo{author}{\bibfnamefont{C.~K.} \bibnamefont{Mishra}},
  \bibinfo{author}{\bibfnamefont{K.~G.} \bibnamefont{Arun}},
  \bibinfo{author}{\bibfnamefont{B.~R.} \bibnamefont{Iyer}}, \bibnamefont{and}
  \bibinfo{author}{\bibfnamefont{B.~S.} \bibnamefont{Sathyaprakash}},
  \bibinfo{journal}{Phys. Rev.} \textbf{\bibinfo{volume}{D82}},
  \bibinfo{pages}{064010} (\bibinfo{year}{2010}), \eprint{1005.0304}.

\bibitem[{\citenamefont{Barack and Cutler}(2004)}]{Barack:2003fp}
\bibinfo{author}{\bibfnamefont{L.}~\bibnamefont{Barack}} \bibnamefont{and}
  \bibinfo{author}{\bibfnamefont{C.}~\bibnamefont{Cutler}},
  \bibinfo{journal}{Phys. Rev.} \textbf{\bibinfo{volume}{D69}},
  \bibinfo{pages}{082005} (\bibinfo{year}{2004}), \eprint{gr-qc/0310125}.

\bibitem[{\citenamefont{{Hild} et~al.}(2011)\citenamefont{{Hild}, {Abernathy},
  {Acernese}, {Amaro-Seoane}, {Andersson}, {Arun}, {Barone}, {Barr},
  {Barsuglia}, {Beker} et~al.}}]{2011CQGra..28i4013H}
\bibinfo{author}{\bibfnamefont{S.}~\bibnamefont{{Hild}}},
  \bibinfo{author}{\bibfnamefont{M.}~\bibnamefont{{Abernathy}}},
  \bibinfo{author}{\bibfnamefont{F.}~\bibnamefont{{Acernese}}},
  \bibinfo{author}{\bibfnamefont{P.}~\bibnamefont{{Amaro-Seoane}}},
  \bibinfo{author}{\bibfnamefont{N.}~\bibnamefont{{Andersson}}},
  \bibinfo{author}{\bibfnamefont{K.}~\bibnamefont{{Arun}}},
  \bibinfo{author}{\bibfnamefont{F.}~\bibnamefont{{Barone}}},
  \bibinfo{author}{\bibfnamefont{B.}~\bibnamefont{{Barr}}},
  \bibinfo{author}{\bibfnamefont{M.}~\bibnamefont{{Barsuglia}}},
  \bibinfo{author}{\bibfnamefont{M.}~\bibnamefont{{Beker}}},
  \bibnamefont{et~al.}, \bibinfo{journal}{Classical and Quantum Gravity}
  \textbf{\bibinfo{volume}{28}}, \bibinfo{pages}{094013}
  (\bibinfo{year}{2011}), \eprint{1012.0908}.

\bibitem[{Sat()}]{Sathya}
\emph{\bibinfo{title}{{\rm Data provided by B.~Sathyaprakash}}}.

\bibitem[{Ber()}]{Berti}
\emph{\bibinfo{title}{{\rm Data provided by E.~Berti}}}.

\bibitem[{\citenamefont{Berti et~al.}(2005{\natexlab{b}})\citenamefont{Berti,
  Buonanno, and Will}}]{Berti:2005qd}
\bibinfo{author}{\bibfnamefont{E.}~\bibnamefont{Berti}},
  \bibinfo{author}{\bibfnamefont{A.}~\bibnamefont{Buonanno}}, \bibnamefont{and}
  \bibinfo{author}{\bibfnamefont{C.~M.} \bibnamefont{Will}},
  \bibinfo{journal}{Class. Quant. Grav.} \textbf{\bibinfo{volume}{22}},
  \bibinfo{pages}{S943} (\bibinfo{year}{2005}{\natexlab{b}}),
  \eprint{gr-qc/0504017}.

\bibitem[{\citenamefont{Stavridis and Will}(2009)}]{Stavridis:2009mb}
\bibinfo{author}{\bibfnamefont{A.}~\bibnamefont{Stavridis}} \bibnamefont{and}
  \bibinfo{author}{\bibfnamefont{C.~M.} \bibnamefont{Will}},
  \bibinfo{journal}{Phys. Rev.} \textbf{\bibinfo{volume}{D80}},
  \bibinfo{pages}{044002} (\bibinfo{year}{2009}), \eprint{0906.3602}.

\bibitem[{\citenamefont{Keppel and Ajith}(2010)}]{Keppel:2010qu}
\bibinfo{author}{\bibfnamefont{D.}~\bibnamefont{Keppel}} \bibnamefont{and}
  \bibinfo{author}{\bibfnamefont{P.}~\bibnamefont{Ajith}},
  \bibinfo{journal}{Phys. Rev.} \textbf{\bibinfo{volume}{D82}},
  \bibinfo{pages}{122001} (\bibinfo{year}{2010}), \eprint{1004.0284}.

\bibitem[{\citenamefont{Yagi and Tanaka}(2009)}]{Yagi:2009zm}
\bibinfo{author}{\bibfnamefont{K.}~\bibnamefont{Yagi}} \bibnamefont{and}
  \bibinfo{author}{\bibfnamefont{T.}~\bibnamefont{Tanaka}}
  (\bibinfo{year}{2009}), \eprint{0906.4269}.

\end{thebibliography}
\end{document}